\documentclass[preprint,12pt,number]{elsarticle}

\usepackage[utf8]{inputenc}
\usepackage[T1]{fontenc}
\usepackage{lmodern}
\usepackage{amsmath,amssymb,amsfonts}
\usepackage{empheq}
\usepackage{algorithm}
\usepackage{algorithmic}
\usepackage{enumitem}
\usepackage{siunitx}
\usepackage{bm}
\usepackage{graphicx}
\usepackage{caption}
\usepackage{subcaption}
\usepackage{booktabs}
\usepackage[section]{placeins}
\usepackage{microtype}
\usepackage[hidelinks]{hyperref}
\renewcommand{\eqref}[1]{Eq.~\textup{(\ref{#1})}}
\newcommand{\eqsref}[2]{Eqs.~\textup{(\ref{#1})} and~\textup{(\ref{#2})}}
\usepackage{xcolor}
\newcommand{\regI}{Regime~I}
\newcommand{\regII}{Regime~II}
\usepackage{setspace}
\usepackage{geometry}
\newcommand{\Div}[1]{\text{div}\left(#1\right)}
\newcommand{\Grad}{\nabla}
\newcommand{\xx}{\mathbf{x}}
\newcommand{\XX}{\mathbf{X}}
\newcommand{\Id}{\mathbf{I}}
\newcommand{\BB}{\mathbf{B}}
\newcommand{\FF}{\mathbf{F}}
\newcommand{\LL}{\mathbf{L}}
\newcommand{\uu}{\mathbf{u}}
\newcommand{\us}{\mathbf{u}_s}
\newcommand{\uf}{\mathbf{u}_f}
\newcommand{\sigg}{\boldsymbol{\sigma}}
\newcommand{\sigs}{\boldsymbol{\sigma}_s}
\newcommand{\sigf}{\boldsymbol{\sigma}_f}

\newcommand{\Tr}{\mathrm{tr}}
\newcommand{\DD}{\mathbf{D}}
\newcommand{\II}{\boldsymbol{I}}
\newcommand{\half}{\tfrac{1}{2}}

\newcommand{\Ftot}{\mathbf{f}}
\newcommand{\dx}{\Delta x}
\newcommand{\dt}{\Delta t}
\newcommand{\Order}[1]{\mathcal{O}\!\left(#1\right)}
\newcommand{\cs}{c_s}

\DeclareMathOperator{\minmod}{minmod}

\newcommand{\dd}{\,\mathrm{d}}

\begin{document}

\begin{frontmatter}

\title{A Monolithic Fully Eulerian Lattice Boltzmann Framework\\
for Fluid--Structure Interaction with Neo-Hookean Elastic Solids}

\author[lrc]{Tristan Millet\corref{cor1}}
\ead{tristan.millet@univ-lr.fr}
\author[lrc]{Erwan Liberge}
\ead{erwan.liberge@univ-lr.fr}

\cortext[cor1]{Corresponding author}
\affiliation[lrc]{organization={LaSIE -- UMR CNRS 7356, Université de La Rochelle},
              addressline={Avenue Michel Crépeau},
              city={La Rochelle},
              postcode={17042},
              country={France}}

\begin{abstract}
\noindent
We present a fully Eulerian, monolithic lattice Boltzmann method for
fluid--structure interaction with incompressible neo-Hookean solids at
unequal densities, and use it to compute the Turek--Hron FSI2 and FSI3
benchmarks on a single Cartesian grid. A pressure-based kinetic scheme
advances the mixture momentum, a polynomial MTHINC reconstruction transports
the solid volume fraction with closed-form fluxes, and the left
Cauchy--Green tensor gives the elastic stress; all updates are explicit and
local. Because the density is transported independently of the populations,
two corrections are introduced. A lattice-link correction makes uniform
translation an exact discrete solution for any density ratio, which is
verified to round-off up to $\rho_s/\rho_f=1000$. A pressure-mode
correction, derived from an asymptotic analysis of the scheme, removes the
frame-dependent defect that otherwise causes a secular drift of the solid
mass. The verification separates the two sources of error: the scheme is
second-order accurate towards the diffuse-interface solution at fixed
interface thickness, and first-order towards the sharp-interface solution
when the thickness scales with the grid, as predicted by an estimate of the
interfacial truncation error. The method reproduces a deformable disk in
Taylor--Green and lid-driven-cavity flows, and the FSI3 and FSI2 benchmarks
at density ratios of one and ten.
\end{abstract}

\end{frontmatter}

\section{Introduction}
\label{sec:intro}
Fluid--structure interaction (FSI) involving large deformations and the
transport of soft solids arises in biomechanics, soft robotics, and
suspensions of deformable particles. In arbitrary Lagrangian--Eulerian
formulations, a body-fitted fluid mesh follows the moving interface, which
may require remeshing when mesh distortions become large. Many
immersed-boundary and fictitious-domain formulations retain a fixed fluid
grid while representing the solid on a separate Lagrangian
mesh~\cite{zhao_fixed-mesh_2008,roy_benchmarking_2015}. These approaches
require the transfer of kinematic and dynamic quantities between the fluid
and solid discretisations. Fully Eulerian formulations provide an
alternative in which both phases and the solid deformation are represented
on a fixed spatial grid, and they avoid the transfer of quantities between two solvers.

In a fully Eulerian monolithic formulation, a common velocity field and a
single set of balance laws describe the fluid and the solid. The term
monolithic refers here to a single-field formulation for both fluid and solid. An indicator
function distinguishes the phases, while an advected Eulerian field records
the solid deformation. This field may be the displacement or initial
position~\cite{dunne_eulerian_2006,cottet_eulerian_2008,deborde_numerical_2020},
the reference map~\cite{kamrin2012,valkov_eulerian_2015,rycroft_reference_2020},
or the left Cauchy--Green
tensor~\cite{sugiyama_full_2011,ii_implicit_2011,ii_full_2012,esmailzadeh_numerical_2014,nishiguchi_full_2019,shin_interaction_2020}.
The tensor formulation used by Sugiyama et
al.~\cite{sugiyama_full_2011} in a volume-of-fluid framework gives the
neo-Hookean stress directly through an upper-convected transport equation.
Its accuracy depends on the transport of both the interface and the
deformation, and on the consistency of their coupling to the momentum
balance. Jain et al.~\cite{jain_conservative_2019} addressed these issues
through a conservative, non-dissipative discretisation, while Prajapati et
al.~\cite{prajapati_monolithic_2025} investigated a geometric
volume-of-fluid reconstruction based on piecewise linear interfaces.
These developments motivate separate assessments of interface transport,
solid incompressibility, and the errors introduced by the interfacial
representation.
The lattice Boltzmann method (LBM)~\cite{kruger_lattice_2017} offers a natural
basis for an explicit formulation on a Cartesian grid: collision is local,
streaming connects neighbouring lattice nodes, and the pressure update does
not require a global Poisson solve. In FSI, it is mostly used for the fluid
alone, coupled through an immersed-boundary method to a Lagrangian solid
solver, typically finite elements~\cite{cai_2019,li_coupling_2019,fringand_2024}.
These partitioned approaches require transfer operators between the two
discretisations, and explicit couplings call for dedicated stabilisation at
density ratios close to one~\cite{fringand_2024}.

Fully Eulerian monolithic LB formulations, in which a single set of
populations describes both phases, remain few. Liberge and
B\'eghein~\cite{liberge_lattice_2022} incorporate the neo-Hookean stress
through a modified equilibrium, track the phases with a Cahn--Hilliard
equation and compute the displacement by finite differences. Sun and
Rycroft~\cite{sun_fully-integrated_2025} couple the reference map technique
to a single-relaxation-time LB solver: a smooth flux correction relaxes the
kinetic density towards its prescribed value, and the reference map is
extrapolated by least squares, layer by layer, over a zone wider than the
transition region to evaluate the solid stress there. Wu et
al.~\cite{wu_fully_2026} combine a conservative Allen--Cahn LB equation for
the interface, a pressure-based multiple-relaxation-time LB equation for the
mixture momentum and the transport of the left Cauchy--Green tensor; the
mobility of the phase-field equation, which balances interface
regularisation against numerical diffusion and stability, is made
time-dependent and set for each case. This formulation is the closest to the
present one.

These formulations have been assessed on neutrally or nearly neutrally
buoyant solids, with density ratios up to three
in~\cite{sun_fully-integrated_2025} and matched or nearly matched densities
in~\cite{wu_fully_2026}, and on flows that drive the solid: lid-driven
cavity, Taylor--Green vortex, settling, channel flow past a deformable block.
To our knowledge, none has addressed the Turek--Hron
benchmarks~\cite{turek2006}, in which a self-sustained flutter grows from an
instability of the coupled system, nor any fluid--structure configuration at
a density ratio of ten.

The obstacle is structural. When the density is prescribed by an interface
scheme rather than carried by the populations, each population arriving at
a node carries the density of the node it left, not that of the node it
reaches. Across a transition resolved by a few grid points, this mismatch
breaks even uniform translation and makes the solid gain or lose mass in a
way that depends on the frame of reference. The present work identifies
these two defects and corrects them, which makes the Turek--Hron FSI2 and
FSI3 benchmarks accessible to a fully Eulerian lattice Boltzmann method for
the first time. The contributions are the following.
\begin{itemize}[leftmargin=*,itemsep=1pt]
  \item A lattice-link form of the density-gradient sources
  of~\cite{liang_phase-field-based_2014}: each population receives the
  equilibrium share of the density displaced along its link. Uniform
  translation becomes an exact discrete solution for any density ratio,
  whereas a Taylor expansion of the same correction, or the flux correction
  of~\cite{sun_fully-integrated_2025}, loses the velocity of the solid.
  \item An asymptotic analysis of the pressure equation that splits the
  kinetic mass error of the solid into two end-point terms, which remain
  bounded, and a frame-dependent defect, which drifts. A correction along the
  pressure mode removes this defect without changing momentum or viscous
  stress.
  \item A verification that separates the convergence towards the diffuse-
  and sharp-interface solutions, together with an estimate of the
  interfacial truncation error, of order $\dx^2/\varepsilon$, which accounts
  for the reduced rates observed when the interface spans a fixed number of
  grid points.
\end{itemize}
As in~\cite{wu_fully_2026}, the solid deformation is represented by the left
Cauchy--Green tensor, transported with the lattice velocity, so that the
algorithm requires neither a Lagrangian solid mesh nor the extrapolation of a
reference map outside the solid. The interface is captured by the algebraic
volume-of-fluid MTHINC reconstruction~\cite{ii_interface_2012}, implemented
with a cubic profile whose fluxes are integrated in closed form; unlike the
phase-field equations of~\cite{liberge_lattice_2022,wu_fully_2026}, it
involves no mobility parameter.

Section~\ref{sec:formulation} presents the continuous model and
Section~\ref{sec:numerics} the numerical method. Section~\ref{sec:validation}
reports the verification tests and Section~\ref{sec:benchmarks} the
fluid--structure benchmarks: a deformable disk in a Taylor--Green
vortex~\cite{zhao_fixed-mesh_2008} and in a lid-driven
cavity~\cite{sugiyama_full_2011}, and the Turek--Hron FSI2 and FSI3
configurations~\cite{turek2006}, at solid-to-fluid density ratios of ten and
one; Section~\ref{sec:conclusion} concludes. The
appendices give the discretisation details and the asymptotic analysis.

\section{Eulerian fluid--solid model}
\label{sec:formulation}
\label{sec:mixture}

\begin{figure}[h]
\centering
\includegraphics{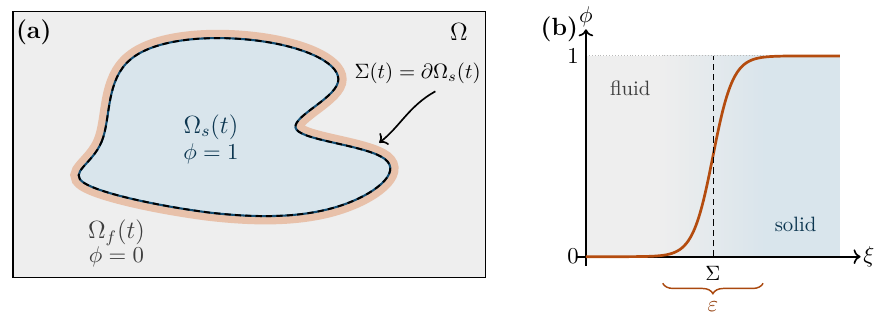}

\caption{Monolithic Eulerian setting.
(a)~Fixed computational domain
$\Omega=\Omega_s(t)\cup\Omega_f(t)$: the solid subdomain $\Omega_s(t)$
($\phi=1$) is embedded in the fluid $\Omega_f(t)$ ($\phi=0$), separated by
the interface $\Sigma(t)=\partial\Omega_s(t)$ (dashed) which is regularised
over a thin diffuse layer of thickness $\varepsilon$ (shaded band).
(b)~Corresponding profile of the volume fraction $\phi$ along the interface
normal $\xi$.}
\label{fig:domain}
\end{figure}

We consider a fixed computational domain $\Omega\subset\mathbb{R}^d$
($d=2$ or $3$), occupied by an incompressible Newtonian fluid and an
incompressible neo-Hookean solid. At each time $t\ge0$, the solid occupies
the subdomain $\Omega_s(t)$, while the fluid occupies the complementary
region $\Omega_f(t)=\Omega\setminus\overline{\Omega_s(t)}$. Their common
boundary is denoted by
$\Sigma(t)=\partial\Omega_s(t)$ (Figure~\ref{fig:domain}).

Instead of tracking the interface explicitly, we introduce the solid volume
fraction $\phi(\xx,t)\in[0,1]$, equal to $1$ in the solid and to $0$ in the
fluid, the transition being regularised over a thin diffuse layer of
thickness $\varepsilon$. The material properties are interpolated through the
mixture laws
\begin{subequations}
\label{eq:mixture_avg}
\begin{empheq}[left=\empheqlbrace\,]{align}
\rho(\phi) &= \phi\rho_s+(1-\phi)\rho_f,
\\[2pt]
\mu(\phi) &= \rho(\phi)\,\nu,
\\[2pt]
G(\phi) &= \phi G_s,
\end{empheq}
\end{subequations}
where $G_s$ is the shear modulus of the neo-Hookean solid; the fluid carries
no elasticity. The kinematic viscosity $\nu$ is uniform, so that
$\mu_f=\rho_f\nu$, $\mu_s=\rho_s\nu$ and $\mu(\phi)$ is the arithmetic
average of $\mu_f$ and $\mu_s$. This law is produced by a uniform relaxation
time of the kinetic scheme without any additional force
(Section~\ref{sec:trt}).

The elastic state of the solid is measured by the left Cauchy--Green tensor
\begin{equation}
\label{eq:def_B}
\BB=\FF\FF^{T},
\qquad
\FF=\frac{\partial\xx}{\partial\XX},
\end{equation}
where $\FF$ is the deformation gradient mapping a material point $\XX$ of the
reference configuration onto its current position $\xx=\xx(\XX,t)$.

The governing equations are solved over the whole computational domain
$\Omega$. They consist of the incompressibility constraint, the momentum
balance, the transport equation of $\BB$, and the advection equation of the
volume fraction,
\begin{subequations}
\label{eq:full_system}
\begin{empheq}[left=\empheqlbrace\,]{align}
\Div\uu &= 0,
\label{eq:full_div}
\\[3pt]
\frac{\partial (\rho \uu ) }{\partial t}
+\Div{\rho\,\uu\otimes\uu}
&=
-\Grad p
+\Div{\sigg(\phi)},
\label{eq:full_mom}
\\[3pt]
\frac{\partial\BB}{\partial t}
+
(\uu\cdot\Grad)\BB
&=
\LL\BB+\BB\LL^{T},
\label{eq:full_B}
\\[3pt]
\frac{\partial\phi}{\partial t}
+\uu\cdot\Grad\phi
&=
0.
\label{eq:full_phi}
\end{empheq}
\end{subequations}

Here,
\[
\DD(\uu)=
\frac12\left(
\Grad\uu+\Grad\uu^T
\right),
\qquad
\LL=\Grad\uu.
\]
\eqref{eq:full_B} follows from $D\FF/Dt=\LL\FF$ along
material trajectories and from $\BB=\FF\FF^T$~\cite{marsden1994}. It implies
$D(\det\BB)/Dt=2\det\BB\,\Div\uu$, so that $\det\BB=1$ is preserved along
trajectories where $\Div\uu=0$.
The non-pressure Cauchy stress is the phase interpolation of
$\sigf=2\mu_f\DD(\uu)$ and $\sigs=2\mu_s\DD(\uu)+G_s(\BB-\Id)$,
\begin{equation}
  \sigg(\phi)
  =2\mu(\phi)\DD(\uu)
  +\phi G_s\left(\BB-\Id\right).
  \label{eq:mixture_stress_expanded}
\end{equation}

The single-field system~\eqref{eq:full_system} enforces the fluid--structure
coupling implicitly. As shown by Mokbel et al.~\cite{mokbel_phase-field_2018},
it converges towards the corresponding sharp-interface problem in the limit
$\varepsilon\rightarrow0$, recovering the kinematic and dynamic transmission
conditions on $\Sigma(t)$,
\begin{subequations}
\label{eq:transmission}
\begin{empheq}[left=\empheqlbrace\,]{align}
\uf &= \us,
\label{eq:transmission_u}
\\[3pt]
\left(-p\,\II+\sigf\right)\mathbf{n}
&=
\left(-p\,\II+\sigs\right)\mathbf{n},
\label{eq:transmission_t}
\end{empheq}
\end{subequations}
where $\mathbf{n}$ is the unit normal to $\Sigma(t)$, and $\uf$, $\us$ are
the traces of the velocity on the fluid and solid sides.
\section{Numerical method}
\label{sec:numerics}
\label{sec:grid}

\subsection{Overview of the time step}
\label{sec:overview}
The system~\eqref{eq:full_system} is advanced on a single uniform Cartesian
lattice of spacing $\dx$ and time step $\dt$, with one velocity field for the
whole domain. All fields are collocated at the
nodes $\xx_{ij}=(i\dx,j\dx)$: the lattice Boltzmann populations
$\{f_q\}_{q=0}^{8}$, which carry the momentum and the pressure, the volume
fraction $\phi$, which sets the density through~\eqref{eq:mixture_avg}, and
the tensor $\BB$. Each field has its own discretisation: a pressure-based
kinetic scheme for the mixture momentum (Section~\ref{sec:lbm}), with two
corrections for the transported density (Sections~\ref{sec:link}
and~\ref{sec:pressure_mode}); a polynomial MTHINC scheme for $\phi$
(Section~\ref{sec:mthinc_impl}); and an upwind finite-difference scheme for
$\BB$ (Section~\ref{sec:muscl_impl}), from which the elastic force is
assembled (Section~\ref{sec:stress_impl}). The last two are advected by the
lattice velocity, and the elastic force enters the collision as a body force. Algorithm~\ref{alg:main} gives the order of the updates.

\begin{algorithm}[!htbp]
\caption{One time step $t^n\rightarrow t^{n+1}$ of the monolithic solver}
\label{alg:main}
\begin{algorithmic}[1]
\REQUIRE $\{f_q^n\}$, $\uu^n$, $p^n$, $\phi^n$, $\BB^n$, $\Ftot^n$ and
$\mathcal R^{n-1}$
\STATE Advect $\phi^n$ with the polynomial MTHINC scheme and
       the velocity $\uu^n$, and set $\rho^{n+1}=\rho(\phi^{n+1})$.
\STATE Apply the source-free TRT collision~\eqref{eq:trt} with
       $f_q^{\mathrm{eq}}(\rho^n,\uu^n,p^n)$.
\STATE
Regularise the non-equilibrium part
       with~\eqref{eq:reg_projector}.
\STATE
Add the forcing
       $\mathcal S_q(\Ftot^n)$ and the link
       correction~\eqref{eq:link_correction} to obtain
       $f_q^{\mathrm{post}}$ from~\eqref{eq:lbm_post}.
\STATE Stream $f_q^{\mathrm{post}}$ along $\mathbf c_q$ and impose the boundary
       conditions.
\STATE Reconstruct $\uu^{n+1}$ from~\eqref{eq:macro_u} with $\Ftot^n$, and the
       provisional pressure $p^\star$ from~\eqref{eq:macro_p}.
\STATE Apply the pressure-mode correction~\eqref{eq:pressure_projection_main}
       to obtain $p^{n+1}$.
\STATE Advance $\BB^n$ with MC--MUSCL--AB2 and the velocity $\uu^{n+1}$, and
       apply the fluid reset~\eqref{eq:B_reset} with $\phi^{n+1}$.
\STATE Assemble $\boldsymbol{\sigma}_{\mathrm{el}}^{n+1}$ and
       $\Ftot^{n+1}$ from~\eqref{eq:div_sigma}, for the next step.
\end{algorithmic}
\end{algorithm}

\subsection{Pressure-based kinetic scheme}
\label{sec:lbm}

\subsubsection{Pressure-based equilibrium}
\label{sec:ilbm}
The D2Q9 lattice Boltzmann method evolves nine populations
$\{f_q(\xx,t)\}_{q=0}^{8}$ on a uniform Cartesian lattice. The D2Q9
velocities are
\begin{equation}
  \mathbf{c}_q=c\,
  \begin{cases}
    (0,0), & q=0,\\
    (1,0),(0,1),(-1,0),(0,-1), & q=1,\ldots,4,\\
    (1,1),(-1,1),(-1,-1),(1,-1), & q=5,\ldots,8,
  \end{cases}
  \qquad c=\frac{\dx}{\dt},
  \label{eq:ci}
\end{equation}
with weights $\omega_0=4/9$, $\omega_{1\ldots4}=1/9$ and
$\omega_{5\ldots8}=1/36$. The lattice sound speed is
$\cs^2=c^2/3$, and $\bar q$ denotes the direction opposite to $q$,
$\mathbf{c}_{\bar q}=-\mathbf{c}_q$. Each time step consists of a local
collision followed by streaming,
\begin{equation}
  f_q(\xx+\mathbf{c}_q\dt,t+\dt)=f_q^{\mathrm{post}}(\xx,t),
  \label{eq:lbe}
\end{equation}
where the post-collision state $f_q^{\mathrm{post}}$ is defined in
Section~\ref{sec:trt}.

We use the pressure-based formulation of Liang et
al.~\cite{liang_phase-field-based_2014}. The mixture density is not a kinetic
variable: it is prescribed by the transported volume fraction
through~\eqref{eq:mixture_avg}, whereas the populations carry the momentum
$\rho\uu$ and the hydrodynamic pressure $p$. The equilibrium reads
\begin{equation}
  f_q^{\mathrm{eq}}(\rho,\uu,p)=a_q\,p+\rho\,\Gamma_q(\uu),
  \qquad
  a_q=\frac{\omega_q-\delta_{q0}}{\cs^2},
  \label{eq:feq}
\end{equation}
with
\begin{equation}
  \Gamma_q(\uu)=\omega_q\left[
    \frac{\mathbf{c}_q\cdot\uu}{\cs^2}
    +\frac{(\mathbf{c}_q\cdot\uu)^2}{2\cs^4}
    -\frac{|\uu|^2}{2\cs^2}
  \right].
  \label{eq:gamma}
\end{equation}
Its moments are
\begin{equation}
  \sum_q f_q^{\mathrm{eq}}=0,
  \qquad
  \sum_q f_q^{\mathrm{eq}}\mathbf{c}_q=\rho\uu,
  \qquad
  \sum_q f_q^{\mathrm{eq}}\mathbf{c}_q \otimes  \mathbf{c}_q
  =p\,\Id+\rho\,\uu \otimes \uu.
  \label{eq:moments}
\end{equation}

\subsubsection{TRT collision and second-order forcing}
\label{sec:trt}

For each opposite pair $(q,\bar q)$, we define
\begin{equation*}
  f_q^{\pm}=\frac{1}{2}(f_q\pm f_{\bar q}),
  \qquad
  f_q^{\mathrm{eq},\pm}=\frac{1}{2}(f_q^{\mathrm{eq}}\pm f_{\bar q}^{\mathrm{eq}}),
  \qquad
  \mathcal{S}_q^{\pm}=\frac{1}{2}(\mathcal{S}_q\pm\mathcal{S}_{\bar q}).
\end{equation*}
The source-free two-relaxation-time (TRT)
collision~\cite{ginzburg_TRT} is
\begin{equation}
  f_q^{\mathrm{c}}=f_q
  -\lambda_+\bigl(f_q^+-f_q^{\mathrm{eq},+}\bigr)
  -\lambda_-\bigl(f_q^--f_q^{\mathrm{eq},-}\bigr),
  \label{eq:trt}
\end{equation}
where $\lambda_\pm=1/\tau_\pm$. The even relaxation time determines the kinematic viscosity,
\begin{equation}
  \nu=\cs^2\left(\tau_+-\frac{1}{2}\right)\dt,
  \label{eq:lbm_viscosity}
\end{equation}
whereas $\tau_-$ is fixed through the TRT parameter
\begin{equation}
  \Lambda=\left(\tau_+-\frac{1}{2}\right)
           \left(\tau_--\frac{1}{2}\right).
  \label{eq:magic}
\end{equation}
Since the populations carry momentum, the forcing is built from the force per unit volume $\Ftot$ itself. We use the
second-order Hermite representation~\cite{GUO2002}
\begin{equation}
  \mathcal{S}_q=\omega_q\left[
    \frac{\mathbf{c}_q\cdot\Ftot}{\cs^2}
    +\frac{(\mathbf{c}_q\cdot\uu)(\mathbf{c}_q\cdot\Ftot)}{\cs^4}
    -\frac{\uu\cdot\Ftot}{\cs^2}
  \right],
  \label{eq:guo}
\end{equation}
whose moments are
\begin{equation}
  \sum_q\mathcal{S}_q=0,
  \qquad
  \sum_q \mathcal{S}_q \mathbf{c}_q=\Ftot,
  \qquad
  \sum_q  \mathcal{S}_q \mathbf{c}_q \otimes  \mathbf{c}_q
  =\uu \otimes  \Ftot+\Ftot \otimes  \uu.
  \label{eq:guo_moments}
\end{equation}
Although its total zeroth moment vanishes, the rest component
$\mathcal{S}_0=-\omega_0\,\uu\cdot\Ftot/\cs^2$ does not, so that the forcing
contributes to the moving-population moment from which the pressure is
recovered (Section~\ref{sec:reconstruction}).

The post-collision populations are finally
\begin{equation}
  f_q^{\mathrm{post}}
  =\mathcal{R}\bigl[f^{\mathrm{c}}\bigr]_q
  +\dt\left(1-\frac{\lambda_+}{2}\right)\mathcal{S}_q^+
  +\dt\left(1-\frac{\lambda_-}{2}\right)\mathcal{S}_q^-
  +\Delta_q,
  \label{eq:lbm_post}
\end{equation}
where $\mathcal{R}$ is the kinetic regularisation of
Section~\ref{sec:regularization} and $\Delta_q$ is the link correction of
Section~\ref{sec:link}. The regularisation acts on the source-free TRT state
only, so that neither the forcing nor the link correction is altered by it.

The body force $\Ftot$ is the
elastic force
\begin{equation}
  \Ftot=\Div{\phi G_s\left(\BB-\Id\right)},
  \label{eq:Ftot}
\end{equation}
to which external body forces, such as gravity, would be added; none is
present in the computations of this paper.
\subsubsection{Macroscopic reconstruction}
\label{sec:reconstruction}
After streaming and boundary treatment, the momentum is reconstructed with
the usual trapezoidal correction,
\begin{equation}
  \rho\uu=\sum_q f_q\mathbf{c}_q+\frac{\dt}{2}\Ftot,
  \label{eq:macro_u}
\end{equation}
where $\rho=\rho(\phi^{n+1})$ is the density delivered by the phase
transport. The pressure is recovered from the moving-population moment
$m=\sum_{q\neq0}f_q$. Inverting the equilibrium while accounting for the
forcing carried by the rest population gives
\begin{equation}
  p=\frac{\cs^2}{1-\omega_0}\left[
      m+\rho\,\Gamma_0(\uu)
      +\frac{\dt}{2}\sum_{q\neq0}\mathcal{S}_q
    \right]
   =\frac{\cs^2}{1-\omega_0}m
    -\frac{\omega_0}{2(1-\omega_0)}\rho|\uu|^2
    +\frac{\omega_0}{2(1-\omega_0)}\dt\,\uu\cdot\Ftot.
  \label{eq:macro_p}
\end{equation}
The last term removes the dependence of the pressure equation on the
relaxation rate and on the work of $\Ftot$ (\ref{app:order4}).

\subsubsection{Kinetic regularisation}
\label{sec:regularization}

The source-free post-collision state is regularised before the forcing and
the link correction are added. Let
$f_q^{\mathrm{neq}}=f_q^{\mathrm{c}}-f_q^{\mathrm{eq}}$, and define its moving
moments
\begin{equation}
  M_{\mathrm{n}}=\sum_{q\neq0}f_q^{\mathrm{neq}},
  \qquad
  \mathbf{j}_{\mathrm{n}}=\sum_{q\neq0}\mathbf{c}_qf_q^{\mathrm{neq}},
  \qquad
  T_{\mathrm{n}}=\sum_{q\neq0}|\mathbf{c}_q|^2f_q^{\mathrm{neq}}
  \label{eq:reg_moments}
\end{equation}
(lattice units, $c=1$). Among the eight moving populations, the five moments
$M_{\mathrm{n}}$, $\mathbf{j}_{\mathrm{n}}$, $\Pi_{xx}-\Pi_{yy}$ and
$\Pi_{xy}$ are kept, and the three others, two odd ghost modes and the
isotropic even mode $T_{\mathrm{n}}-\tfrac65M_{\mathrm{n}}$, are set to their
equilibrium value. The operation is therefore equivalent to a
multiple-relaxation-time collision~\cite{lallemand_theory_2000}
with unit relaxation rate for these three modes; such collisions are also used in
pressure-based multiphase schemes~\cite{liang_phase-field-based_2014}. The equilibrium of the isotropic mode is
then fixed by the ratio of the moving weights,
$\sum_{q\neq0}|\mathbf{c}_q|^2\omega_q/\sum_{q\neq0}\omega_q=6/5$, and the
projector reads
\begin{equation}
  (\mathcal{P}f^{\mathrm{neq}})_q
  =f_q^{\mathrm{neq},+}
  +3\omega_q\,\mathbf{c}_q\cdot\mathbf{j}_{\mathrm{n}}
  +\eta_q\left(\frac{6}{5}M_{\mathrm{n}}-T_{\mathrm{n}}\right),
  \quad q\neq0,
  \qquad
  (\mathcal{P}f^{\mathrm{neq}})_0=f_0^{\mathrm{neq}},
  \label{eq:reg_projector}
\end{equation}
with $\eta_q=-1/4$ for the axial and $+1/4$ for the diagonal directions; the
regularised state is
$\mathcal{R}[f^{\mathrm{c}}]_q=f_q^{\mathrm{eq}}+(\mathcal{P}f^{\mathrm{neq}})_q$.
Since $M_{\mathrm{n}}$, $\mathbf{j}_{\mathrm{n}}$ and the deviatoric stress
are preserved, the incompressibility constraint, the shear
viscosity~\eqref{eq:lbm_viscosity} and the momentum equation are unchanged at
the orders retained in~\ref{app:asymptotics}. The odd ghost modes no longer relax with $\lambda_-$, so that the TRT
relation~\eqref{eq:magic} only sets the relaxation of the non-equilibrium
momentum. Applying $\mathcal{P}$ to the link increment would break
property~\eqref{eq:translation_exact}, which is why $\Delta_q$ is added
afterwards.

\subsection{Link correction}
\label{sec:link}
The density is supplied by the phase transport, not by a kinetic moment.
Without correction, a population reaching $\xx+\mathbf{c}_q\dt$ carries the
density $\rho^n(\xx)$ of its departure node, while~\eqref{eq:macro_u} uses
$\rho^{n+1}(\xx+\mathbf{c}_q\dt)$: a uniform translation is not preserved in
the interfacial layer, and the moving scalar moment yields $\Div{\rho\uu}=0$
instead of $\Div\uu=0$ (\ref{app:order12}). Liang et
al.~\cite{liang_phase-field-based_2014} compensate these errors with
finite-difference density-gradient sources and a term
$\tfrac12\dt\,\uu\cdot\Grad\rho$ in the pressure closure. We use their
lattice counterpart: each population receives the equilibrium share of the
density displaced along its link,
\begin{equation}
  \Delta_q^n(\xx)
  =\Gamma_q\bigl(\uu_{\ell,q}^n\bigr)
   \left[\rho^{n+1}(\xx+\mathbf{c}_q\dt)-\rho^n(\xx)\right],
  \label{eq:link_correction}
\end{equation}
with the link velocity $\uu_{\ell,q}^n$ the mass-weighted average of $\uu^n$
over the nodes of the cell crossed by the link,
\begin{equation}
  \begin{aligned}
  \uu_{\ell,q}^n
  &=\frac{\sum_{\mathbf{y}\in\mathcal C_q(\xx)}\rho^n(\mathbf{y})\,\uu^n(\mathbf{y})}
        {\sum_{\mathbf{y}\in\mathcal C_q(\xx)}\rho^n(\mathbf{y})},
  \\[4pt]
  \mathcal C_q(\xx)
  &=
  \begin{cases}
    \bigl\{\xx,\ \xx+\mathbf{c}_q\dt\bigr\} & \text{axial links},\\[2pt]
    \bigl\{\xx,\ \xx+c_{qx}\mathbf{e}_x\dt,\ \xx+c_{qy}\mathbf{e}_y\dt,\
    \xx+\mathbf{c}_q\dt\bigr\} & \text{diagonal links}.
  \end{cases}
  \end{aligned}
  \label{eq:link_velocity}
\end{equation}
For smooth fields, the leading moments of $\Delta_q$ are the sources
of~\cite{liang_phase-field-based_2014} (\ref{app:LC_expansion}), with
$\partial_t\rho$ replaced by the increment actually delivered by the phase
transport. Since $\Delta_q$ is streamed with its population, no half-step
term is needed in~\eqref{eq:macro_p}.

Uniform translation becomes exact. In a periodic domain without body force,
let $f_q^n=f_q^{\mathrm{eq}}(\rho^n,\mathbf{U},p_0)$. The collision leaves
$f^n$ unchanged, the regularisation $\mathcal{R}$ acts on a vanishing
non-equilibrium part, $\uu_{\ell,q}=\mathbf{U}$ and, $f^{\mathrm{eq}}$ being
affine in $\rho$,
$f_q^{\mathrm{post}}(\xx)=f_q^{\mathrm{eq}}\bigl(\rho^{n+1}(\xx+\mathbf{c}_q\dt),\mathbf{U},p_0\bigr)$.
After streaming, \eqsref{eq:macro_u}{eq:macro_p} return $\mathbf{U}$ and
$p_0$, the correction of Section~\ref{sec:pressure_mode} vanishes and, by
induction,
\begin{equation}
  f_q^{n+1}=f_q^{\mathrm{eq}}(\rho^{n+1},\mathbf{U},p_0).
  \label{eq:translation_exact}
\end{equation}

\subsection{Pressure-mode correction}
\label{sec:pressure_mode}
In a region of uniform density, the expansion of~\ref{app:order4} gives
\begin{equation}
  \frac{1-\omega_0}{\rho\cs^2}
  \left(\frac{Dp}{Dt}-\mathcal{N}_p\right)
  +\Div\uu=\Order{Ma^4},
  \label{eq:pressure_eulerian_main}
\end{equation}
with
\begin{equation}
  \mathcal{N}_p
  =\frac95\,\uu\cdot\Grad p
  +\frac45\,\rho\,\uu\cdot\Grad K
  -\frac45\,\uu\cdot\Div{2\rho\nu\DD(\uu)},
  \qquad K=\frac12|\uu|^2.
  \label{eq:Np}
\end{equation}
The material term $Dp/Dt$ is the Galilean-invariant artificial-compressibility
error. The defect $\mathcal{N}_p$ depends on the velocity of the frame: it
comes from the moving-population closure and has no physical counterpart.

The coefficient of $Dp/Dt$ sets the artificial sound speed
$c_a=\cs/\sqrt{1-\omega_0}$, identical in both phases.\footnote{In the solid,
the bulk modulus $\rho_s c_a^2\dx^2/\dt^2$ is compared with $G_s$ through the
elastic Mach number
$Ma_{\mathrm{el}}=c_a^{-1}\sqrt{G_s/\rho_s}\,\dt/\dx$; the solid is
incompressible only for $Ma_{\mathrm{el}}\to0$. Since
$\dt=(\tau_+-\tfrac12)\dx^2/(3\nu)$,
$Ma_{\mathrm{el}}\propto(\tau_+-\tfrac12)\dx$.}

Where the density varies, $\mathcal{N}_p$ becomes $\mathcal{N}_p+\mathcal{N}_\rho$
(\ref{app:order4}), with
\begin{equation}
  \mathcal{N}_\rho
  =-\frac{3}{10}\,\partial_t\uu\cdot\Grad\rho
  +\frac{3}{20}\,\cs^2\,\nabla^2\uu\cdot\Grad\rho
  +\frac{3}{10}\,\frac{\cs^2}{\rho}\,\Grad\rho\cdot\DD(\uu)\cdot\Grad\rho,
  \label{eq:Nrho}
\end{equation}
which vanishes for uniform $\rho$ and in uniform translation.

After streaming and reconstruction of $\uu^\star$ and $p^\star$, the defect
is removed along the pressure mode,
\begin{equation}
  \delta_p=\dt\,\mathcal{N}_p(\uu^\star,p^\star),
  \qquad
  f_q\leftarrow f_q-a_q\,\delta_p,
  \qquad
  p\leftarrow p^\star-\delta_p,
  \label{eq:pressure_projection_main}
\end{equation}
with $a_q$ from~\eqref{eq:feq}. Since $a_q$ has zero zeroth and first
moments and unit isotropic second moment, momentum and velocity are
unchanged and the non-equilibrium viscous stress is not reset. Only
$\mathcal{N}_p$ is projected: it is the only kinetic contribution to the mass
of the solid that can produce a secular drift (\ref{app:mass}). The
truncation error of the phase transport and the interfacial layer remain, so
that the mass of the solid is not conserved exactly. Without the correction,
a fast-moving solid erodes its volume fraction and loses the
incompressibility of its deformation (Section~\ref{sec:turek},
Figure~\ref{fig:turek_fsi3_galilean}).

\subsection{Interface capturing: polynomial MTHINC}
\label{sec:mthinc_impl}
The discrete value $\phi_{ij}$ denotes the cell average of the solid volume
fraction. Its material transport equation is written in conservative form as
\begin{equation}
  \frac{\partial\phi}{\partial t}+\Div{\phi\uu}
  =\phi\Div\uu,
  \label{eq:phi_conservative}
\end{equation}
where the right-hand side retains the small residual dilatation of the
pressure-based lattice Boltzmann solver.

The reconstruction follows the multidimensional THINC
scheme~\cite{ii_interface_2012}, with the hyperbolic tangent replaced by a
cubic smoothed Heaviside function. The
construction is referred to as the polynomial MTHINC scheme. Integrating~\eqref{eq:phi_conservative} over a cell and one time step gives
\begin{equation}
  \phi_{ij}^{n+1}=\phi_{ij}^{n}
  -\left(F_{i+\half,j}^{\phi}-F_{i-\half,j}^{\phi}\right)
  -\left(F_{i,j+\half}^{\phi}-F_{i,j-\half}^{\phi}\right)
  +S_{ij}^{\phi},
  \label{eq:phi_fv}
\end{equation}
where $F^{\phi}$ denotes the time-integrated swept-volume flux and
$S_{ij}^{\phi}$ the discretisation of $\phi\Div\uu$. The face velocity is the
two-point average of the adjacent cell-centred velocities, smoothed in the
transverse direction with the weights $(1,4,1)/6$,
\begin{equation}
  u_{i+\half,j}=\frac1{12}\Bigl[
    4\left(u_{ij}+u_{i+1,j}\right)
    +u_{i,j-1}+u_{i+1,j-1}+u_{i,j+1}+u_{i+1,j+1}
  \Bigr],
  \label{eq:mthinc_face_avg}
\end{equation}
and analogously for $v_{i,j+\half}$. The difference of two consecutive face
velocities reproduces exactly the isotropic derivative~\eqref{eq:lbm_grad},
\begin{equation}
  \frac{u_{i+\half,j}-u_{i-\half,j}}{\dx}=\mathcal D_xu\big|_{ij},
  \label{eq:face_div}
\end{equation}
so that the dilatation seen by the volume fraction is the one used for the
transport of $\BB$ and for the elastic force (Sections~\ref{sec:muscl_impl}
and~\ref{sec:stress_impl}). Inside an interfacial
cell, the interface is represented by
\begin{equation}
  \mathcal H_{ij}(X,Y)
  =H\!\left(\beta\,[n_xX+n_yY+d]\right),
  \qquad (X,Y)\in[0,1]^2,
  \label{eq:mthinc_reconstruction_main}
\end{equation}
where $H$ is the cubic smoothed Heaviside, equal to
$\tfrac12+\tfrac34s-\tfrac14s^3$ for $|s|<1$, to $0$ for $s\le-1$ and to $1$
for $s\ge1$, $(n_x,n_y)$ is the Youngs normal, and $d$ is chosen so that the
cell average of $\mathcal H_{ij}$ equals $\phi_{ij}$. Since $(X,Y)$ is
measured in cell units, the profile rises from $0$ to $1$ over the thickness
\begin{equation}
  \varepsilon=\frac{2\dx}{\beta},
  \label{eq:eps_beta}
\end{equation}
and is exactly $0$ or $1$ outside this layer. The flux is the integral of~\eqref{eq:mthinc_reconstruction_main}
over the strip swept during the time step. All required one- and
two-dimensional primitives, the safeguarded solution for $d$, and the
axis-aligned limiting cases are given in \ref{app:mthinc}.

The two-dimensional update is performed by alternating directional sweeps. The
source $\phi\Div\uu$ is treated by the alternating implicit/explicit rule
detailed in \ref{app:mthinc_time}. For an $x$-first step,
\begin{equation}
  \phi_{ij}^{*}
  =\frac{\phi_{ij}^{n}
  -\left(F_{i+\half,j}^{\phi}-F_{i-\half,j}^{\phi}\right)}
  {1-\dt\,(\partial_xu)_{ij}},
  \qquad
  \phi_{ij}^{n+1}
  =\phi_{ij}^{*}\left[1+\dt\,(\partial_yv)_{ij}\right]
  -\left(F_{i,j+\half}^{\phi}-F_{i,j-\half}^{\phi}\right),
  \label{eq:thinc_sweep}
\end{equation}
with the order exchanged on the following step, where
$(\partial_xu)_{ij}$ and $(\partial_yv)_{ij}$ are the face
differences~\eqref{eq:face_div}. Pure cells, $\phi_{ij}\le\phi_c$ or
$\phi_{ij}\ge1-\phi_c$ with $\phi_c=10^{-8}$, bypass the reconstruction, while
interfacial cells use the closed-form polynomial integrals exclusively.

The MTHINC flux does not by itself keep $\phi$ in $[0,1]$ once the
dilatation source is included. Each sweep is therefore closed by a
flux-corrected transport limiter~\cite{zalesak_fully_1979}, with the
first-order upwind update as the bounded low-order solution
(\ref{app:mthinc_flux}); a single flux is retained on each shared face, so
that the limiter preserves the discrete conservation of $\phi$.

\subsection{Transport of the left Cauchy--Green tensor}
\label{sec:muscl_impl}
The deformation is transported through the upper-convected
equation, \eqref{eq:full_B}. In two dimensions, the three independent components
$(B_{xx},B_{xy},B_{yy})$ are advanced on the same collocated grid. The
advection term is discretised with a MUSCL reconstruction and the Monotonized-Central limiter applied
independently to each component; the upwind face state is selected by the sign
of the nodal velocity (\ref{app:muscl}).

The velocity gradient $\LL$ is evaluated by the centred isotropic D2Q9
operator used for the elastic-stress divergence. The complete semi-discrete
right-hand side is
\begin{equation}
  \mathcal R(\BB,\uu)
  =-(\uu\cdot\Grad)_{\mathrm{MC\text{-}MUSCL}}\BB
   +\LL\BB+\BB\LL^T.
  \label{eq:rhs_B}
\end{equation}
After two forward-Euler start-up steps, the tensor is advanced using the second-order Adams--Bashforth (AB2) scheme:
\begin{equation}
  \BB^{n+1}=
  \begin{cases}
    \BB^n+\dt\,\mathcal R^n, & n\in\{0,1\},\\[3pt]
    \BB^n+\dt\left(\dfrac32\mathcal R^n
                    -\dfrac12\mathcal R^{n-1}\right), & n\ge2,
  \end{cases}
  \qquad
  \mathcal R^n=\mathcal R(\BB^n,\uu^{n+1}),
  \label{eq:ab2_B}
\end{equation}
where $\uu^{n+1}$ is the velocity delivered by the kinetic update of the same
step (Algorithm~\ref{alg:main}).

Solid incompressibility implies $\det\BB=1$ along material trajectories. We
monitor $\det\BB-1$ as a numerical diagnostic. Since~\eqref{eq:full_B} is
solved on the whole fixed grid, tensor values in cells that have become pure
fluid are reset to the stress-free state,
\begin{equation}
  \BB_{ij}^{n+1}=\Id
  \qquad\text{when}\qquad \phi_{ij}^{n+1}<\phi_{\mathrm{thr}},
  \label{eq:B_reset}
\end{equation}
which prevents deformation stored in the fluid region from re-entering the
solid through numerical advection. The threshold is the pure-cell tolerance of
the interface scheme, $\phi_{\mathrm{thr}}=\phi_c=10^{-8}$, in all
computations, so that the rule introduces no additional parameter.
\subsection{Elastic stress and body-force assembly}
\label{sec:stress_impl}
The transported fields $(\phi^{n+1},\BB^{n+1})$ determine the elastic
stress
\begin{equation}
  \boldsymbol{\sigma}_{\mathrm{el},ij}
  =\phi_{ij}G_s\left(\BB_{ij}-\Id\right).
  \label{eq:sigma_cell}
\end{equation}
Its divergence is evaluated with the fourth-order-isotropic D2Q9 stencil. For
a scalar cell field $q$,
\begin{equation}
  \mathcal D_xq\big|_{ij}
  =\frac{1}{12\dx}\left[
    4(q_{i+1,j}-q_{i-1,j})
    +q_{i+1,j+1}+q_{i+1,j-1}
    -q_{i-1,j+1}-q_{i-1,j-1}
  \right],
  \label{eq:lbm_grad}
\end{equation}
with $\mathcal D_y$ obtained by exchanging $x$ and $y$. Hence
\begin{equation}
  \Ftot_x=\mathcal D_x\sigma_{\mathrm{el},xx}
             +\mathcal D_y\sigma_{\mathrm{el},xy},
  \qquad
  \Ftot_y=\mathcal D_x\sigma_{\mathrm{el},xy}
             +\mathcal D_y\sigma_{\mathrm{el},yy}.
  \label{eq:div_sigma}
\end{equation}

The force~\eqref{eq:Ftot}, evaluated with~\eqref{eq:div_sigma}, enters the
forcing term~\eqref{eq:guo}, the momentum reconstruction~\eqref{eq:macro_u}
and the pressure reconstruction~\eqref{eq:macro_p} of the next step.
\subsection{Implementation}
\label{sec:cost}
The method is implemented in C++17 and parallelised with OpenMP. On a 20-core
Intel Xeon w5-3535X processor, the FSI2 computations of
Section~\ref{sec:turek}, all steps of Algorithm~\ref{alg:main} included, run
at 34 million lattice-node updates per second with 12 threads on the
$2\,\mathrm{mm}$ grid ($1.5\times10^5$ nodes), and at 73 million with
26 threads on the $1.25\,\mathrm{mm}$ grid ($4.0\times10^5$ nodes), the two
computations sharing the processor.

\section{Verification}
\label{sec:validation}
\label{sec:verif}

All computations use Algorithm~\ref{alg:main} with the same discrete
operators: the isotropic stencil~\eqref{eq:lbm_grad} for $\LL$, for the
elastic force and for the face velocities~\eqref{eq:mthinc_face_avg}, and the
TRT parameter $\Lambda=1/12$ and the regularisation of Section~\ref{sec:regularization}.

Two global indicators monitor the transport of the solid phase. The mass error
\begin{equation}
  \frac{\Delta M}{M_0}(t)
  =\frac{\sum_{ij}\phi_{ij}(t)-\sum_{ij}\phi_{ij}(0)}{M_0}
  \label{eq:mass_error}
\end{equation}
measures the conservation of $\phi$, where $M_0=\sum_{ij}\phi_{ij}(0)$ is
restricted to the deformable part of the body when a rigid obstacle is
present. The mean Jacobian
\begin{equation}
  \bar J(t)=\frac{\sum_{\mathcal S}\phi_{ij}\sqrt{\det\BB_{ij}}}
                 {\sum_{\mathcal S}\phi_{ij}},
  \qquad
  \mathcal S=\{(i,j):\ \phi_{ij}>0.1\},
  \label{eq:mean_J}
\end{equation}
measures the incompressibility of the elastic solid, for which $J=\sqrt{\det\BB}=1$ exactly. Neither quantity is
enforced by the scheme. This section checks, on problems with exact solutions,
the two discrete properties on which the method rests: uniform translation is
exact at any density ratio, and the scheme is second-order accurate for the
diffuse-interface model, so that its convergence towards the sharp-interface
problem is limited by the mixture laws only. The benchmarks of
Section~\ref{sec:benchmarks} then assess the full fluid--structure coupling
against published results.

\subsection{Uniform translation}
A neo-Hookean disk of radius $0.2L$ ($G_s=1$) is translated with the uniform
velocity $\mathbf{U}=(0.01,0.0075)$ (lattice units) across the periodic square
$[0,L]^2$, with $N=128$, $\beta=0.5$, $\tau_+=1$ and the populations
initialised at equilibrium. The exact solution is $\uu\equiv\mathbf{U}$,
$p\equiv p_0$, $\BB\equiv\Id$, which the scheme reproduces to round-off for
any density ratio~\eqref{eq:translation_exact}. Table~\ref{tab:translation}
confirms it up to $\rho_s/\rho_f=1000$ and compares two alternatives: the
second-order Taylor expansion of the link difference
$\rho^{n+1}(\xx+\mathbf{c}_q\dt)-\rho^n(\xx)$ around $\xx$, with centred
differences, which has the same moments up to $\Order{\dx^3}$ but whose
neglected terms are not small across a four-node interface, and the smooth
flux correction of Sun and Rycroft~\cite{sun_fully-integrated_2025}, run
with the authors' code\footnote{Repository \texttt{github.com/yue-sun/lbrmt},
version of February 2025, with two changes: the array of rigid-obstacle flags
is initialised to zero, which the code leaves uninitialised when no obstacle
file is given, and gravity is set to zero.} ($\tau=1$, default transition
half-width of $1.5$ cells), which keeps the density of the solid but not its
momentum.

\begin{table}[!htbp]
  \centering
  \caption{Uniform translation of a disk, $N=128$, $\tau_+=1$: maximum
  relative velocity error $\max|\uu-\mathbf{U}|/|\mathbf{U}|$ and, for the
  link correction, maximum mass error, after one crossing of the domain
  ($12\,800$ steps). Sun and Rycroft: first $4000$ steps, before the
  extrapolation zone of the reference map reaches the periodic boundary; in
  parentheses, mean velocity of the solid core divided by $|\mathbf{U}|$.}
  \label{tab:translation}
  \begin{tabular}{lcccc}
    \toprule
    & \multicolumn{4}{c}{$\rho_s/\rho_f$} \\
    \cmidrule(lr){2-5}
    & $1.5$ & $10$ & $100$ & $1000$ \\
    \midrule
    Link correction~\eqref{eq:link_correction}
      & -- & $1.0\times10^{-12}$ & $1.0\times10^{-12}$ & $1.1\times10^{-12}$ \\
    \quad mass error $\max|\Delta M/M_0|$
      & -- & $3.9\times10^{-14}$ & $4.5\times10^{-14}$ & $3.9\times10^{-14}$ \\
    Taylor expansion of $\Delta_q$
      & -- & $4.6\times10^{-1}$ & diverges & diverges \\
    Sun and Rycroft~\cite{sun_fully-integrated_2025}
      & $0.25$ $(0.84)$ & $3.5$ $(0.42)$ & -- & -- \\
    \bottomrule
  \end{tabular}
\end{table}
\subsection{Oscillating fluid--solid layers}
\label{sec:layers}
\begin{figure}[!htbp]
\centering
\includegraphics{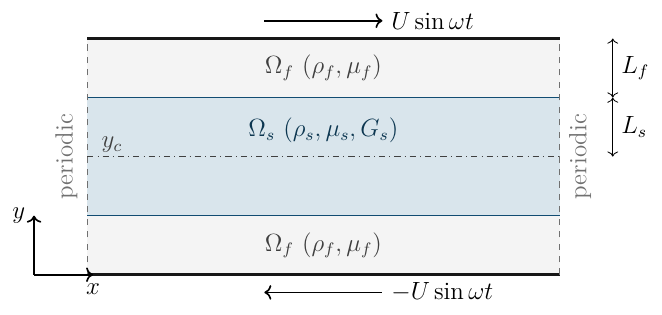}
\caption{Oscillating fluid--solid layers: a neo-Hookean layer between two
fluid layers, driven by walls moving in opposite directions; the domain is
periodic in $x$.}
\label{fig:layers_setup}
\end{figure}
The interface thickness $\varepsilon$ and the MTHINC sharpness $\beta$ are
tied by~\eqref{eq:eps_beta}, $\varepsilon=2\dx/\beta$: $\beta$ sets the
number of nodes across the interfacial layer, $\varepsilon$ its physical
thickness. The error of a diffuse-interface computation with respect to a
sharp-interface solution has two parts. The first is a model error: through
the mixture laws~\eqref{eq:mixture_avg}, which define $\rho$, $\mu$ and $G$
inside the layer, the diffuse-interface problem differs from the
sharp-interface one by $\Order{\varepsilon}$~\cite{sugiyama_full_2011}. The
second is the discretisation error. The analysis of~\ref{app:asymptotics} is
a series in $\dx\,\partial$ and assumes fields that are smooth on the lattice
scale; inside the layer, $\rho$, and wherever $\rho$, $\mu$ or $G$ jumps the
derivatives of the velocity, vary over the length $\varepsilon$, so that the
expansion parameter is $\dx/\varepsilon$. A truncation term of formal order
$\dx^2$ is then $\Order{\dx^2/\varepsilon^2}$ in the layer and, the layer
having the width $\varepsilon$, contributes $\Order{\dx^2/\varepsilon}$ to
integrals over the domain, such as the mass. For such quantities, the error
with respect to the sharp-interface solution is therefore estimated as
\begin{equation}
  \Order{\varepsilon}+\Order{\dx^2/\varepsilon}+\Order{\dx^2},
  \label{eq:interface_estimate}
\end{equation}
where the coefficient of the interfacial term grows with the jumps of
$\rho$, $\mu$ and $G$ across the layer. At uniform density, only the jumps of
$G$, and of $\mu$ if $\nu$ is not uniform, contribute. It defines two
refinement regimes:
\begin{itemize}
  \item \regI{} ($\beta$ fixed, $\varepsilon=\Order{\dx}$): $\dx/\varepsilon$
  is constant, and at unequal densities the density changes by a fixed
  fraction between neighbouring nodes whatever $\dx$. All orders of the
  expansion are of the same size inside the layer, the analysis
  of~\ref{app:asymptotics} does not hold there, and~\eqref{eq:interface_estimate}
  gives first-order convergence towards the sharp-interface problem, which is
  the problem of interest. All the results compared with reference solutions
  are given in this regime.
  \item \regII{} ($\varepsilon$ fixed, $\varepsilon=\Order{1}$):
  $\dx/\varepsilon\to0$, the expansion holds, the model error is frozen and
  the computation converges at second order towards the diffuse-interface
  solution at that thickness. This regime measures the order of the
  discretisation, in particular of the mass conservation.
\end{itemize}
In \regI, a correction derived from the truncated expansion, such as one
based on $\mathcal{N}_\rho$~\eqref{eq:Nrho}, is therefore not expected to
recover second order. The oscillating layers of Sugiyama et
al.~\cite{sugiyama_full_2011} have exact solutions for both regimes and
measure the two parts separately.

A solid layer $|y-y_c|<L_s$ lies between two fluid layers
$L_s<|y-y_c|<L_s+L_f$, bounded by walls at $y-y_c=\pm(L_s+L_f)$ that move with
the velocities $\pm U\sin\omega t$; the domain is periodic in $x$
(Figure~\ref{fig:layers_setup}). As
in~\cite{sugiyama_full_2011}, $L_s=L_f=1/2$, $\rho_f=1$, $\mu_f=1$, $G_s=5$,
$\omega=\pi$ and $U=1$; here $\mu_s=\rho_s\nu$
(Section~\ref{sec:mixture}) instead of $\mu_s=0$, and $\rho_s/\rho_f=1$ and
$10$. The computations use $\tau_+=1$, $N_y$ cells across the channel and
four columns, start from rest, and cover ten periods; the errors are
evaluated over the last period, at $21$ instants. The computed velocity is
independent of $x$ to round-off.

For $\uu=(u(y,t),0)$, the transport of $\BB$ gives $B_{xy}=\gamma$, the shear
strain, and $B_{yy}=1$ exactly, so that the shear stress
$\mu\,\partial_yu+G\gamma$, $G=\phi G_s$, is linear and the problem reduces to
\begin{equation}
  \rho\,\partial_tu=\partial_y\bigl(\mu\,\partial_yu+G\gamma\bigr),
  \qquad
  \partial_t\gamma=\partial_yu .
  \label{eq:layers}
\end{equation}
Two references are used. The sharp-interface solution $u_0$ has piecewise
constant coefficients; in the periodic regime,
$u_0=\mathrm{Re}\,[V(y)\,e^{\mathrm{i}\omega t}]$ with
$(\mu^*V')'=\mathrm{i}\omega\rho V$, $\mu^*=\mu+G/(\mathrm{i}\omega)$, a
combination of hyperbolic functions in each layer, with continuity of $V$ and
$\mu^*V'$ at the
interfaces. The diffuse-interface solution $u_\varepsilon$
solves~\eqref{eq:layers} with $\rho(\phi)$, $\mu(\phi)$ and $G(\phi)$
of~\eqref{eq:mixture_avg} and the initial profile $\phi$ of the computation,
which is not transported since $\uu\cdot\Grad\phi=0$. It is computed from rest
with $4000$ points and the Crank--Nicolson scheme; doubling both resolutions
changes it by less than $3\times10^{-7}$. The start from rest is kept in the reference, because the
strain that it leaves at the edge of the layer, where $\phi G_s$ is small, has
not relaxed over the simulated time.

\begin{figure}[!htbp]
  \centering
  \includegraphics[width=\textwidth]{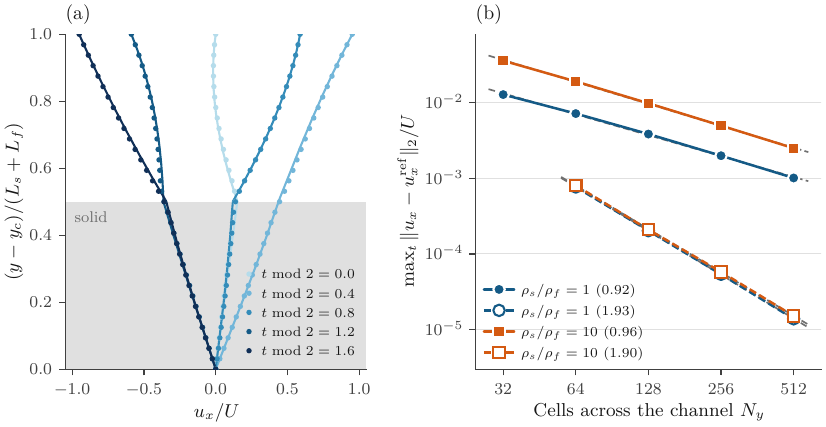}
  \caption{Oscillating fluid--solid layers. (a)~Velocity in the upper half of
  the channel at five instants of the last period, $N_y=64$, $\beta=0.5$,
  $\rho_s/\rho_f=1$ (symbols), and sharp-interface solution (lines).
  (b)~Maximum over the last period of the $L^2$ error: filled symbols,
  $\beta=0.5$ fixed, against the sharp-interface solution; open symbols,
  $\varepsilon=1/8$ fixed, against the diffuse-interface solution. Slopes of
  the least-squares fits in parentheses.}
  \label{fig:layers}
\end{figure}

Figure~\ref{fig:layers}(a) compares the computed profiles with $u_0$, and
Table~\ref{tab:layers} gives the errors. In \regI, the error with respect to
$u_0$ is of first order for both density ratios, and almost all of it is the
model error $\|u_\varepsilon-u_0\|$. In \regII, the error with respect to
$u_\varepsilon$ is of second order in both norms, whereas the error with
respect to $u_0$ saturates at the model error. The scheme is therefore
second-order accurate for the diffuse-interface model, including inside the
layer and across the density jump, and its first-order convergence towards
the sharp-interface solution in \regI{} comes from the mixture laws, as
reported by Sugiyama et al.~\cite{sugiyama_full_2011} for this problem.
\begin{table}[!htbp]
  \centering
  \caption{Oscillating fluid--solid layers: maximum over the last period of
  the $L^2$ error of the velocity. \regI: $\beta=0.5$, error with respect to
  the sharp-interface solution $u_0$. \regII: $\varepsilon=1/8$, error with
  respect to the diffuse-interface solution $u_\varepsilon$. Orders between
  the two finest grids.}
  \label{tab:layers}
  \begin{tabular}{lcccc}
    \toprule
    & \multicolumn{2}{c}{\regI{} ($\beta=0.5$), vs $u_0$}
    & \multicolumn{2}{c}{\regII{} ($\varepsilon=1/8$), vs $u_\varepsilon$} \\
    \cmidrule(lr){2-3}\cmidrule(lr){4-5}
    $N_y$ & $\rho_s/\rho_f=1$ & $\rho_s/\rho_f=10$ & $\rho_s/\rho_f=1$ & $\rho_s/\rho_f=10$ \\
    \midrule
    $32$  & $1.28\times10^{-2}$ & $3.58\times10^{-2}$ & -- & -- \\
    $64$  & $7.17\times10^{-3}$ & $1.90\times10^{-2}$ & $7.67\times10^{-4}$ & $8.04\times10^{-4}$ \\
    $128$ & $3.84\times10^{-3}$ & $9.80\times10^{-3}$ & $2.03\times10^{-4}$ & $2.10\times10^{-4}$ \\
    $256$ & $1.99\times10^{-3}$ & $4.96\times10^{-3}$ & $5.39\times10^{-5}$ & $5.80\times10^{-5}$ \\
    $512$ & $1.01\times10^{-3}$ & $2.50\times10^{-3}$ & $1.39\times10^{-5}$ & $1.52\times10^{-5}$ \\
    \midrule
    Order, $L^2$       & $0.98$ & $0.99$ & $1.96$ & $1.94$ \\
    Order, $L^\infty$  & -- & -- & $2.00$ & $1.98$ \\
    Model error $\|u_\varepsilon-u_0\|$ & \multicolumn{2}{c}{share at $N_y=512$: $90\,\%$ / $\simeq100\,\%$}
      & $6.5\times10^{-3}$ & $1.9\times10^{-2}$ \\
    \bottomrule
  \end{tabular}
\end{table}
In what follows, \regI{} therefore measures the convergence towards the
sharp-interface problem, and \regII{} the accuracy of the discretisation.

\FloatBarrier

\section{Fluid--structure benchmarks}
\label{sec:benchmarks}

Three two-dimensional problems are considered: a disk deformed by a decaying
Taylor--Green vortex (Section~\ref{sec:tgv}), a soft disk transported in a
lid-driven cavity (Section~\ref{sec:ldc2d}), and
the Turek--Hron FSI2 and FSI3 benchmarks~\cite{turek2006}
(Section~\ref{sec:turek}). Without an exact solution, the observed rate of a
quantity $q$ is computed from three successive grids,
\begin{equation}
  r=\log_2\frac{\|q_N-q_{2N}\|}{\|q_{2N}-q_{4N}\|}.
  \label{eq:observed_order}
\end{equation}
Section~\ref{sec:synthesis} collects the observations on the mass and the
Jacobian. Here $N$ is the number of cells along the $x$ or $y$ direction, $q_N$ the quantity computed on the $N^2$ grid.

\subsection{Taylor--Green vortex with an embedded neo-Hookean solid}
\label{sec:tgv}
\begin{figure}[!htbp]
  \centering
  \includegraphics[width=\textwidth]{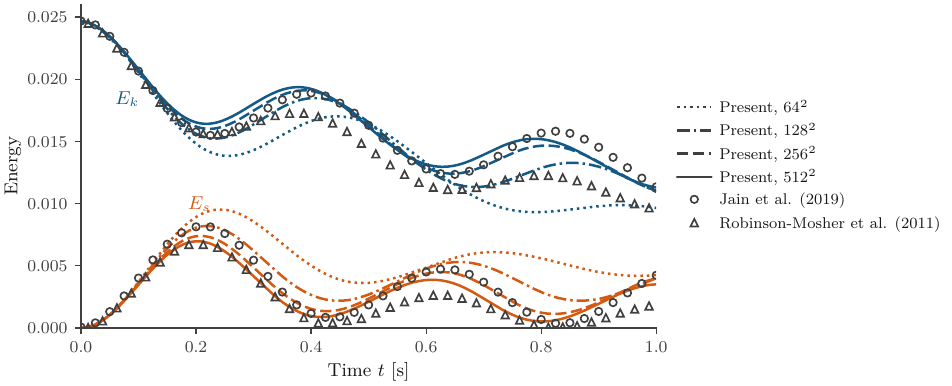}
  \caption{Taylor--Green vortex with an embedded disk, $G_s=1$: kinetic
  energy $E_k$ (blue) and elastic energy $E_s$ (orange) on four grids,
  compared with Jain et al.~\cite{jain_conservative_2019} and
  Robinson-Mosher et al.~\cite{robinson-mosher_symmetric_2011} (as reported
  in~\cite{jain_conservative_2019}).}
  \label{fig:tg_energies}
\end{figure}

A disk of radius $r=0.2L$ is placed at the centre of the periodic
square $[0,L]^2$, $L=1$, and set in motion by the Taylor--Green field
\begin{equation}
  u_x=\psi_0k\sin(kx)\cos(ky),
  \qquad
  u_y=-\psi_0k\cos(kx)\sin(ky),
  \qquad k=2\pi,\quad \psi_0=5\times10^{-2},
  \label{eq:tg_init}
\end{equation}
with $\rho_s=\rho_f=1$, a uniform viscosity $\mu=10^{-3}$ and $G_s=1$, as
in~\cite{zhao_fixed-mesh_2008,robinson-mosher_symmetric_2011,jain_conservative_2019}. The kinetic energy
$E_k=\int_\Omega\tfrac12\rho|\uu|^2\,\dd\xx$ and the elastic energy
$E_s=\int_\Omega\tfrac12\phi G_s(\Tr\BB-2)\,\dd\xx$ are recorded. Four grids, $64^2$ to $512^2$, are used with $\tau_+=0.512$ and
$\beta=0.5$; a second series keeps $\varepsilon=1/32$ fixed ($\beta=0.5$,
$0.25$ and $0.125$ on the $128^2$, $256^2$ and $512^2$ grids) and converges
to the diffuse-interface solution at that thickness, so that the references
are compared with the first series. The phase and period of the energy exchange agree with all references
(Figure~\ref{fig:tg_energies}). Table~\ref{tab:tg} compares the first
extrema of the energies with those of Jain et al.~\cite{jain_conservative_2019},
Robinson-Mosher et al.~\cite{robinson-mosher_symmetric_2011} and Zhao et
al.~\cite{zhao_fixed-mesh_2008}, the last two as reported
in~\cite{jain_conservative_2019}: on the finest grid, the first maximum of
$E_s$ lies within the spread of the references. The mass error decreases at
second order in \regII{} and at a slightly lower rate in \regI. In
\regI, the defect of $\det\BB$ in the interfacial layer reaches the edge of
the bulk; in \regII, it is reduced by more than two orders of magnitude.
\begin{table}[!htbp]
  \centering
  \caption{Taylor--Green vortex with an embedded disk, $\rho_s=\rho_f$:
  first maximum of the elastic energy $E_s$ and first minimum of the kinetic
  energy $E_k$ (time in parentheses), maximum mass
  error~\eqref{eq:mass_error} in \regI{} ($\beta=0.5$) and \regII{}
  ($\varepsilon=1/32$), with the orders between successive grids, and
  maximum of $|\sqrt{\det\BB}-1|$ on the $512^2$ grid. Reference values
  read from Fig.~14(b) of~\cite{jain_conservative_2019}.}
  \label{tab:tg}
  \small
  \begin{tabular}{lcccc}
    \toprule
    & \multicolumn{2}{c}{\regI{} ($\beta=0.5$)} & \multicolumn{2}{c}{$\max|\Delta M/M_0|$} \\
    \cmidrule(lr){2-3}\cmidrule(lr){4-5}
    Grid & $\max E_s\ [10^{-3}]$ & $\min E_k\ [10^{-3}]$ & \regI & \regII \\
    \midrule
    $64^2$  & $9.52$ ($0.242$) & $13.82$ ($0.258$) & $2.5\times10^{-4}$ & -- \\
    $128^2$ & $8.20$ ($0.220$) & $15.22$ ($0.232$) & $7.8\times10^{-5}$ & $7.8\times10^{-5}$ \\
    $256^2$ & $7.39$ ($0.208$) & $16.00$ ($0.222$) & $2.4\times10^{-5}$ & $2.0\times10^{-5}$ \\
    $512^2$ & $6.97$ ($0.202$) & $16.41$ ($0.217$) & $7.3\times10^{-6}$ & $5.0\times10^{-6}$ \\
    Orders  & -- & -- & $1.70$, $1.72$, $1.70$ & $1.98$, $1.97$ \\
    $\max|\sqrt{\det\BB}-1|$, $512^2$ & & & $0.2$ & $9.4\times10^{-4}$ \\
    \midrule
    Jain et al.~\cite{jain_conservative_2019} ($128^2$) & $8.19$ ($0.213$) & $15.46$ ($0.228$) & & \\
    Robinson-Mosher et al.~\cite{robinson-mosher_symmetric_2011} & $6.81$ ($0.200$) & $15.53$ ($0.229$) & & \\
    Zhao et al.~\cite{zhao_fixed-mesh_2008} & $6.75$ ($0.193$) & $16.58$ ($0.229$) & & \\
    \bottomrule
  \end{tabular}
\end{table}

\paragraph{Unequal densities}
With $\rho_s/\rho_f=10$ (Table~\ref{tab:tg_rho10}), the heavier disk
exchanges energy more slowly and deforms more, and its first maximum of
$E_s$ occurs later. In \regII, the mass error is of second order. In \regI,
it is of first order: the interfacial term of~\eqref{eq:interface_estimate}
is $\Order{\dx}$ when $\varepsilon\propto\dx$, and its coefficient grows
with the density jump, which explains why the order is higher at
matched densities (Table~\ref{tab:tg}) and drops to $1$ here.

\begin{table}[!htbp]
  \centering
  \caption{Taylor--Green vortex, $\rho_s/\rho_f=10$, $t\le1$: first maximum
  of the elastic energy, maximum mass error~\eqref{eq:mass_error} and maximum
  deviation of the mean Jacobian~\eqref{eq:mean_J}, with the observed orders
  between successive grids.}
  \label{tab:tg_rho10}
  \begin{tabular}{llccc}
    \toprule
    & Grid & $\max E_s$ & $\max|\Delta M/M_0|$ & $\max|\bar J-1|$ \\
    \midrule
   \regI & $64^2$  & $3.47\times10^{-2}$ & $1.2\times10^{-3}$ & $9.6\times10^{-3}$ \\
                & $128^2$ & $3.30\times10^{-2}$ & $5.6\times10^{-4}$ & $6.6\times10^{-3}$ \\
                & $256^2$ & $3.21\times10^{-2}$ & $2.7\times10^{-4}$ & $3.8\times10^{-3}$ \\
                & $512^2$ & $3.17\times10^{-2}$ & $1.3\times10^{-4}$ & $1.8\times10^{-3}$ \\
                & orders  & -- & $1.09$, $1.02$, $1.03$ & $0.54$, $0.81$, $1.10$ \\
    \midrule
    \regII & $128^2$ & $3.30\times10^{-2}$ & $5.6\times10^{-4}$ & $6.6\times10^{-3}$ \\
                & $256^2$ & $3.28\times10^{-2}$ & $1.4\times10^{-4}$ & $2.9\times10^{-3}$ \\
                & $512^2$ & $3.27\times10^{-2}$ & $3.6\times10^{-5}$ & $6.1\times10^{-4}$ \\
                & orders  & -- & $1.96$, $1.97$ & $1.20$, $2.24$ \\
    \bottomrule
  \end{tabular}
\end{table}

\FloatBarrier
\subsection{Taylor--Green vortex in a moving frame}
\label{sec:tg_frame}
This test checks the pressure-mode correction of
Section~\ref{sec:pressure_mode} on the Taylor--Green case of
Section~\ref{sec:tgv}. Superimposing a uniform velocity $\mathbf{V}=(0,V)$ on~\eqref{eq:tg_init}
only translates the exact solution, so that the mass error should not depend
on $V$. The defect $\mathcal{N}_p$~\eqref{eq:Np} does. With
$\uu=\mathbf{V}+\uu'$, its part quadratic in $\mathbf{V}$ is
$\tfrac45\rho\,\mathbf{V}\cdot\DD(\uu')\cdot\mathbf{V}$, the terms linear in
$\mathbf{V}$ vanish by symmetry, and by~\eqref{eq:app_mass} it changes the
mass of the solid by
\begin{equation}
  \delta_V(t)=\frac43\,V_{\mathrm{lu}}^2\,
  \bar\varepsilon_V(t),
  \qquad
  \bar\varepsilon_V(t)=\int_0^t
  \frac{\int_\Omega\phi\,\mathbf{e}_V\cdot\DD(\uu')\,\mathbf{e}_V\,\dd\xx}
       {\int_\Omega\phi\,\dd\xx}\,\dd s,
  \label{eq:galilean_prediction}
\end{equation}
where $V_{\mathrm{lu}}=V\dt/\dx$ is the superimposed velocity in lattice
units and $\bar\varepsilon_V$ the strain of the solid along
$\mathbf{e}_V=\mathbf{V}/V$, averaged over the solid and integrated in time.
Without the correction~\eqref{eq:pressure_projection_main}, $\delta_V/V_{\mathrm{lu}}^2$
should thus be the same for all velocities and grids.

\begin{figure}[!htbp]
  \centering
  \includegraphics[width=\textwidth]{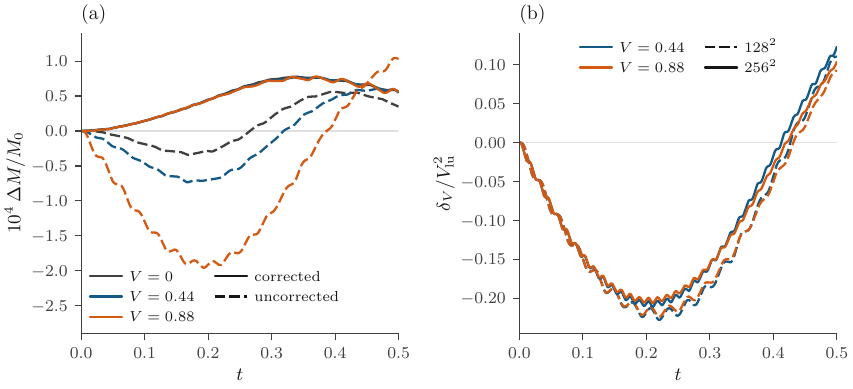}
  \caption{Taylor--Green vortex with a superimposed uniform velocity
  $(0,V)$. (a)~Mass error~\eqref{eq:mass_error} on the $128^2$ grid, with
  (solid) and without (dashed) the pressure-mode
  correction~\eqref{eq:pressure_projection_main}. (b)~Frame-dependent part
  $\delta_V$ of the uncorrected mass error, divided by the square of the
  superimposed velocity in lattice units, on two grids.}
  \label{fig:tg_galilean}
\end{figure}

The runs use $V=0$, $0.4375$ and $0.875$, up to $2.8$ times the maximum
velocity of the vortex, on the $128^2$ and $256^2$ grids with $\beta=0.5$,
for $t\le0.5$, with and without the correction. With the correction, the mass error does not depend on $V$
(Figure~\ref{fig:tg_galilean}(a)). Without it, the frame-dependent part is
measured as
\begin{equation*}
  \delta_V=e_V-e_0,
  \qquad
  e_V=\frac{\Delta M}{M_0}\Big|_{\text{uncorrected}}
     -\frac{\Delta M}{M_0}\Big|_{\text{corrected}}\quad\text{at velocity }V,
\end{equation*}
where $e_V$ isolates the contribution of $\mathcal{N}_p$ and $e_0$ its
frame-independent part. It scales as~\eqref{eq:galilean_prediction} predicts
(Table~\ref{tab:tg_galilean}), and the four curves of
$\delta_V/V_{\mathrm{lu}}^2$ nearly coincide
(Figure~\ref{fig:tg_galilean}(b)).
\begin{table}[!htbp]
  \centering
  \caption{Taylor--Green vortex with a superimposed velocity $(0,V)$,
  $t\le0.5$: frame dependence of the mass error with and without the
  pressure-mode correction~\eqref{eq:pressure_projection_main}.}
  \label{tab:tg_galilean}
  \begin{tabular}{lcc}
    \toprule
    & Predicted~\eqref{eq:galilean_prediction} & Observed \\
    \midrule
    With correction: $\max_{V,t}|\Delta M/M_0(V)-\Delta M/M_0(0)|$ & $0$ & $6\times10^{-6}$ \\
    Without: ratio of $\max_t|\delta_V|$ when $V$ is doubled & $4$ & $3.9$ \\
    Without: ratio of $\max_t|\delta_V|$ when $\dx$ is halved & $4$ & $4.3$ \\
    \bottomrule
  \end{tabular}
\end{table}

\FloatBarrier
\subsection{Neo-Hookean solid in a two-dimensional lid-driven cavity}
\label{sec:ldc2d}
\begin{figure}[!htbp]
\centering
\includegraphics{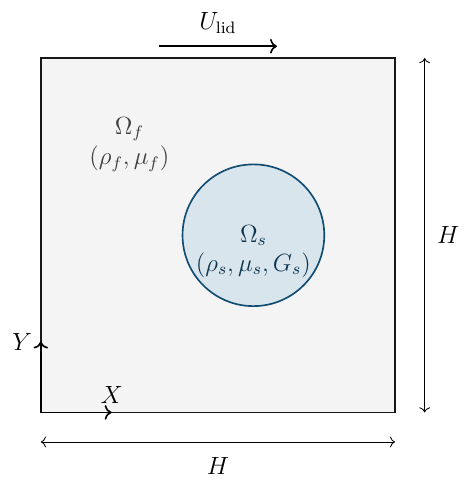}
\caption{Computational domain for the lid-driven cavity with an embedded deformable neo-Hookean solid.}
\label{fig:ldc_setup}
\end{figure}

A neo-Hookean disk of radius $R=0.2H$, initially at rest and
stress-free, is centred at $(0.6H,0.5H)$ in the square cavity $[0,H]^2$
(Figure~\ref{fig:ldc_setup}). The lid translates impulsively at the constant
velocity $U_{\mathrm{lid}}$, and no-slip conditions are imposed on the three
other walls. The four walls are wet nodes, with the velocity imposed by the non-equilibrium
bounce-back of Zou and He~\cite{zou_pressure_1997} and the corner nodes at
equilibrium.  Densities and viscosities are
matched, $\rho_s=\rho_f=1$ and $\mu_s=\mu_f=10^{-2}$, with $H=1$ and
$U_{\mathrm{lid}}=1$. The flow is characterised by the Reynolds and Cauchy
numbers
\begin{equation}
  Re=\frac{\rho_fU_{\mathrm{lid}}H}{\mu_f}=100,
  \qquad
  Ca=\frac{\rho_fU_{\mathrm{lid}}^2}{G_s}.
  \label{eq:ldc_numbers}
\end{equation}
The reference case of~\cite{sugiyama_full_2011} is $G_s=0.1$ ($Ca=10$). The
MTHINC sharpness is $\beta=0.5$, and $\tau_+=1$ in
all runs except $G_s=1000$, for which $\tau_+=0.525$. The simulations cover
$t\in[0,20]$.

\begin{figure}[!htbp]
  \centering
  \includegraphics[width=\textwidth]{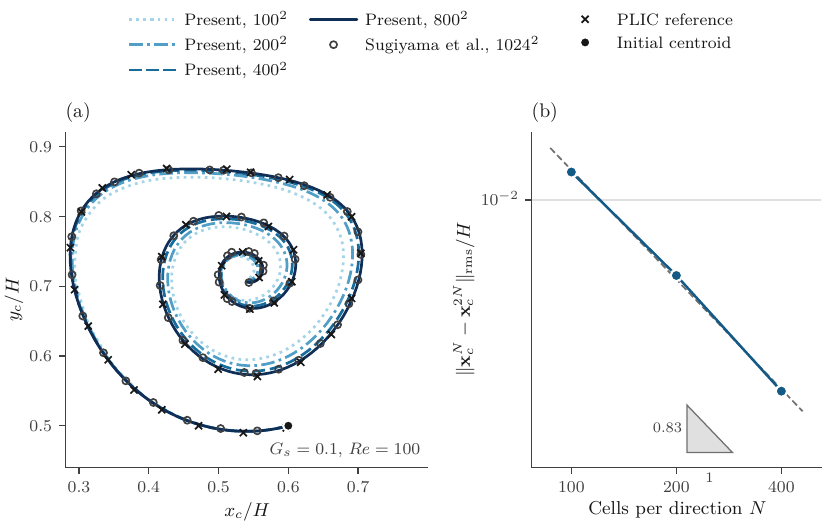}
  \caption{Lid-driven cavity, $Re=100$, $G_s=0.1$. (a)~Centroid trajectories
  over $t\in[0,20]$ on four grids, compared with Sugiyama et
  al.~\cite{sugiyama_full_2011} ($1024^2$) and with the PLIC-based solver of
  Prajapati et al.~\cite{prajapati_monolithic_2025} ($128^2$).
  (b)~Time-matched root-mean-square distance between the centroid
  trajectories on the $N^2$ and $(2N)^2$ grids; the dashed line is a
  least-squares power-law fit.}
  \label{fig:ldc_trajectory_convergence}
\end{figure}

Figure~\ref{fig:ldc_trajectory_convergence}(a) shows the trajectory of the
centroid of the disk for $G_s=0.1$ on four grids, $N^2=100^2$ to $800^2$. The
disk is entrained by the primary vortex and spirals towards its centre. All
resolutions follow the reference trajectories of Sugiyama et
al.\ ($N=1024$)~\cite{sugiyama_full_2011} and of the PLIC-based Eulerian
solver of Prajapati et al.\ ($N=128$)~\cite{prajapati_monolithic_2025}
closely, with observed rates~\eqref{eq:observed_order} close to the
first order expected (Table~\ref{tab:ldc}).

\begin{figure}[!htbp]
  \centering
  \includegraphics[width=0.62\textwidth]{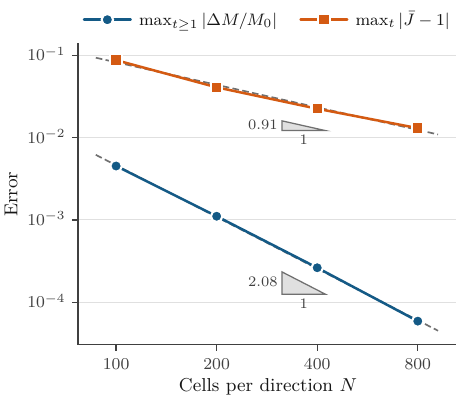}
  \caption{Lid-driven cavity, $Re=100$, $G_s=0.1$: grid convergence of the
  maximum mass error~\eqref{eq:mass_error} over $t\ge1$ and of the maximum
  deviation of the mean Jacobian~\eqref{eq:mean_J}. Dashed lines are
  least-squares power-law fits, with the slopes indicated by the triangles.}
  \label{fig:ldc_conservation_convergence}
\end{figure}

The conservation indicators are reported in
Figure~\ref{fig:ldc_conservation_convergence} and
Table~\ref{tab:ldc}: the mass error decreases at second order, whereas the
mean Jacobian converges at first order.
\begin{table}[!htbp]
  \centering
  \caption{Lid-driven cavity, $Re=100$, $G_s=0.1$, $\beta=0.5$:
  time-matched r.m.s.\ distance between the centroid trajectories on the
  $N^2$ and $(2N)^2$ grids, maximum mass error~\eqref{eq:mass_error} over
  $t\ge1$ and maximum deviation of the mean Jacobian~\eqref{eq:mean_J}.
  Orders: between successive grids for the trajectory, least-squares fit
  over the four grids otherwise.}
  \label{tab:ldc}
  \begin{tabular}{lcccc}
    \toprule
    $N$ & $100$ & $200$ & $400$ & $800$ \\
    \midrule
    Trajectory, $N$ vs $2N$ & $1.2\times10^{-2}$ & $6.7\times10^{-3}$ & $3.7\times10^{-3}$ & -- \\
    \quad order & \multicolumn{3}{c}{$0.78$, $0.87$} & \\
    $\max_{t\ge1}|\Delta M/M_0|$ & $4.5\times10^{-3}$ & $1.1\times10^{-3}$ & $2.6\times10^{-4}$ & $5.9\times10^{-5}$ \\
    \quad order & \multicolumn{4}{c}{$2.08$} \\
    $\max_t|\bar J-1|$ & $8.7\times10^{-2}$ & $4.1\times10^{-2}$ & $2.2\times10^{-2}$ & $1.3\times10^{-2}$ \\
    \quad order & \multicolumn{4}{c}{$0.91$} \\
    \bottomrule
  \end{tabular}
\end{table}

\begin{figure}[!htbp]
  \centering
  \includegraphics[width=0.62\textwidth]{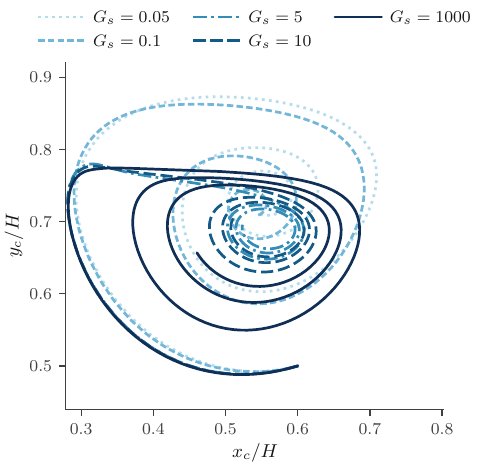}
  \caption{Lid-driven cavity, $Re=100$, $200^2$ grid: centroid trajectories
  over $t\in[0,20]$ for five values of the shear modulus.}
  \label{fig:ldc_shear_modulus}
\end{figure}

\begin{figure}[!htbp]
  \centering
  \includegraphics[width=\textwidth]{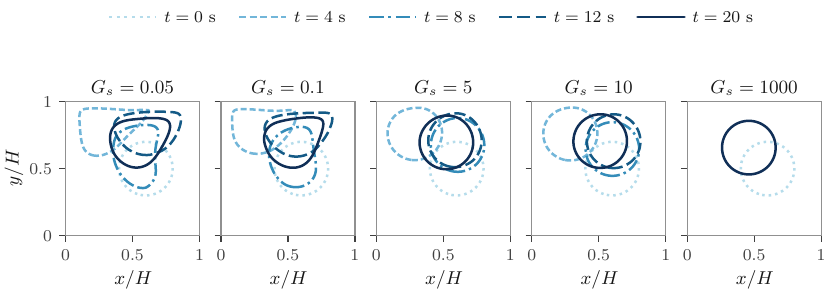}
  \caption{Lid-driven cavity, $Re=100$, $200^2$ grid: contour $\phi=1/2$ of
  the disk at $t=0$, $4$, $8$, $12$ and $20$ for each value of the shear modulus
  ($t=0$ and $20$ only for $G_s=1000$).}
  \label{fig:ldc_shapes}
\end{figure}

The shear modulus is varied over more than four decades,
\begin{equation}
  G_s\in\{0.05,\,0.1,\,5,\,10,\,1000\},
  \qquad
  Ca\in\{20,\,10,\,0.2,\,0.1,\,10^{-3}\},
\end{equation}
on the $200^2$ grid. The centroid trajectories and the shapes of the disk are
shown in Figures~\ref{fig:ldc_shear_modulus} and~\ref{fig:ldc_shapes}. Only $\tau_+$ is changed for $G_s=1000$: the explicit elastic coupling remains stable over the whole range of $Ca$. Table~\ref{tab:ldc_gs} gives the conservation errors over the whole range.
\begin{table}[!htbp]
  \centering
  \caption{Lid-driven cavity, $Re=100$, $200^2$ grid, $\beta=0.5$:
  mass error~\eqref{eq:mass_error} at $t=20$ and maximum deviation of the
  mean Jacobian~\eqref{eq:mean_J} for five shear moduli.}
  \label{tab:ldc_gs}
  \begin{tabular}{ccccc}
    \toprule
    $G_s$ & $Ca$ & $\tau_+$ & $\Delta M/M_0$ ($t=20$) & $\max_t|\bar J-1|$ \\
    \midrule
    $0.05$ & $20$      & $1$     & $9.3\times10^{-4}$ & $4.1\times10^{-2}$ \\
    $0.1$  & $10$      & $1$     & $9.1\times10^{-4}$ & $4.1\times10^{-2}$ \\
    $5$    & $0.2$     & $1$     & $6.7\times10^{-4}$ & $1.3\times10^{-2}$ \\
    $10$   & $0.1$     & $1$     & $3.8\times10^{-4}$ & $8.1\times10^{-3}$ \\
    $1000$ & $10^{-3}$ & $0.525$ & $9.5\times10^{-5}$ & $8.1\times10^{-4}$ \\
    \bottomrule
  \end{tabular}
\end{table}

\FloatBarrier
\subsection{Turek--Hron FSI2 and FSI3 benchmarks}
\label{sec:turek}
\begin{figure}[!htbp]
\centering
\includegraphics{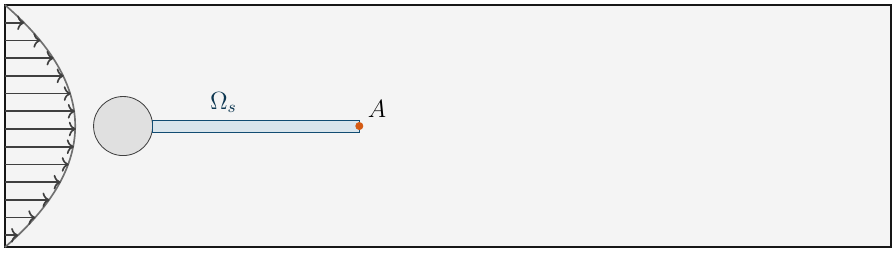}
\caption{Turek--Hron FSI benchmark: elastic bar behind a rigid cylinder with Poiseuille inflow. Point $A$ denotes the midpoint of the free end of the elastic structure.}
\label{fig:turek_poiseuille}
\end{figure}

\begin{figure}[!htbp]
  \centering
  \includegraphics[width=\textwidth]{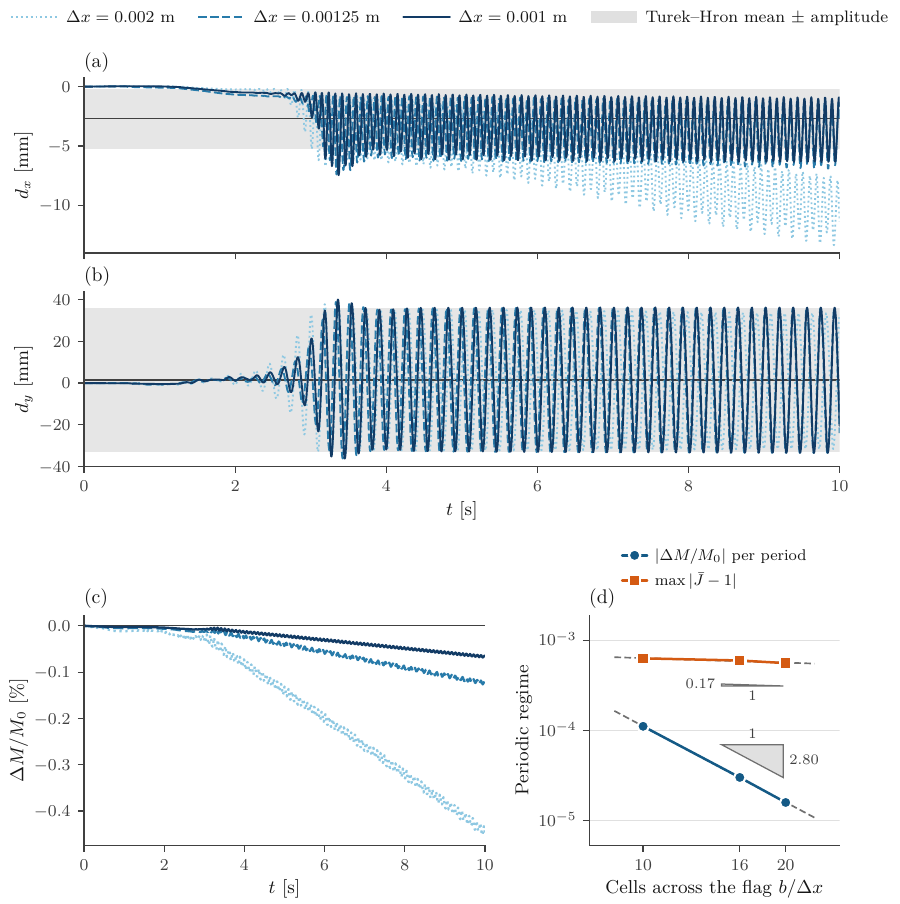}
  \caption{Turek--Hron FSI3, $\beta=0.5$, $\tau_+=0.51$, on three grids:
  (a)~horizontal and (b)~vertical displacement of point $A$, with the range
  mean $\pm$ amplitude (band) and the mean value (line) of the
  reference~\cite{turek2006}; (c)~mass error~\eqref{eq:mass_error} of the
  bar; (d)~last five periods: change of mass per period and maximum
  deviation of the mean Jacobian~\eqref{eq:mean_J}, against the number of
  cells across the bar. The slopes indicated are observed rates.}
  \label{fig:turek_fsi3}
\end{figure}

\begin{figure}[!htbp]
  \centering
  \includegraphics[width=\textwidth]{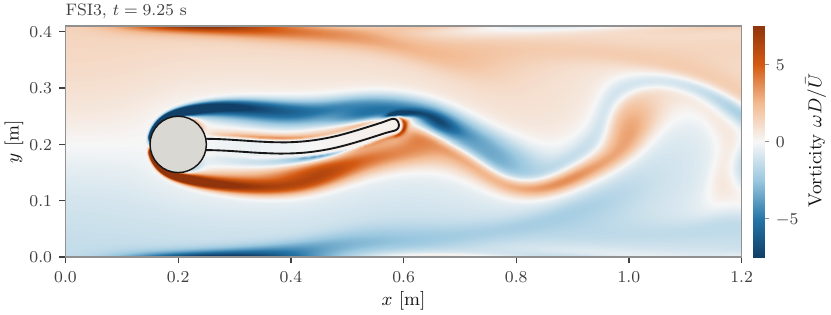}
  \caption{Turek--Hron FSI3, $\dx=2\,\mathrm{mm}$, $\beta=0.5$,
  $\tau_+=0.51$: vorticity $\omega D/\bar U$ in the periodic regime, at
  $t=9.25\,\mathrm{s}$, with the contour $\phi=1/2$ of the bar.}
  \label{fig:turek_fsi3_snapshot}
\end{figure}

\begin{figure}[!htbp]
  \centering
  \includegraphics[width=\textwidth]{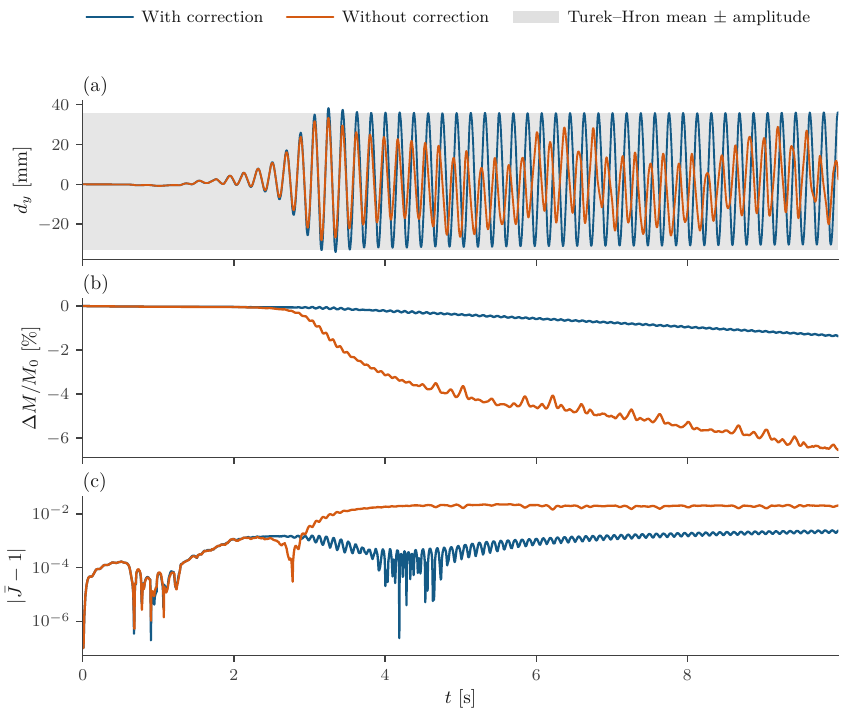}
  \caption{Turek--Hron FSI3, $\dx=2\,\mathrm{mm}$, $\beta=0.5$,
  $\tau_+=0.52$, with and without the pressure-mode
  correction~\eqref{eq:pressure_projection_main}:
  (a)~vertical displacement of point $A$, with the range mean $\pm$
  amplitude of the reference~\cite{turek2006}; (b)~mass
  error~\eqref{eq:mass_error} of the bar; (c)~deviation of the mean
  Jacobian~\eqref{eq:mean_J}.}
  \label{fig:turek_fsi3_galilean}
\end{figure}

The Turek--Hron benchmarks~\cite{turek2006} couple a channel flow with an
elastic bar clamped behind a rigid cylinder (Figure~\ref{fig:turek_poiseuille},
Table~\ref{tab:turek_params}): FSI2 has a heavy, soft bar at $Re=100$, FSI3 a
light, stiffer bar at $Re=200$. The fluid is initially at rest, with no-slip walls, the pressure outlet
of~\cite{junk2011} is used and the inflow
\begin{equation}
  u_{\mathrm{in}} = \frac32\bar U\,\frac{4\,y\,(H-y)}{H^2}\;r(t),
  \qquad
  r(t)=
  \begin{cases}
    \tfrac12\bigl[1-\cos(\pi t/t_r)\bigr], & t<t_r,\\
    1, & t\ge t_r,
  \end{cases}
  \qquad t_r=2\,\mathrm{s},
  \label{eq:turek_inlet}
\end{equation}
with $H=0.41\,\mathrm{m}$. The cylinder, of diameter $D=0.1\,\mathrm{m}$, is treated with a Brinkman
penalisation~\cite{liberge_vp_2019}. Unlike the reference compressible
Saint-Venant--Kirchhoff solid ($\nu_s=0.4$), the bar is incompressible
neo-Hookean with the same shear modulus, and carries the viscosity
$\mu_s=\rho_s\nu_f$.
\paragraph{FSI3}
For
FSI3, the displacement of point $A$ approaches the reference under refinement
(Figure~\ref{fig:turek_fsi3}, Table~\ref{tab:turek_tip}); Figure~\ref{fig:turek_fsi3_snapshot} shows
the flow in the periodic regime. The mean horizontal displacement is the most
sensitive quantity: the erosion of the free end moves point $A$ upstream, by
an amount that decreases with $\dx$. Without the pressure-mode correction
($\tau_+=0.52$, Figure~\ref{fig:turek_fsi3_galilean}), the mass loss is
multiplied several times, the defect of $\bar J$ grows by one order of
magnitude, and the oscillation of the bar loses amplitude and regularity.

\begin{figure}[!htbp]
  \centering
  \includegraphics[width=\textwidth]{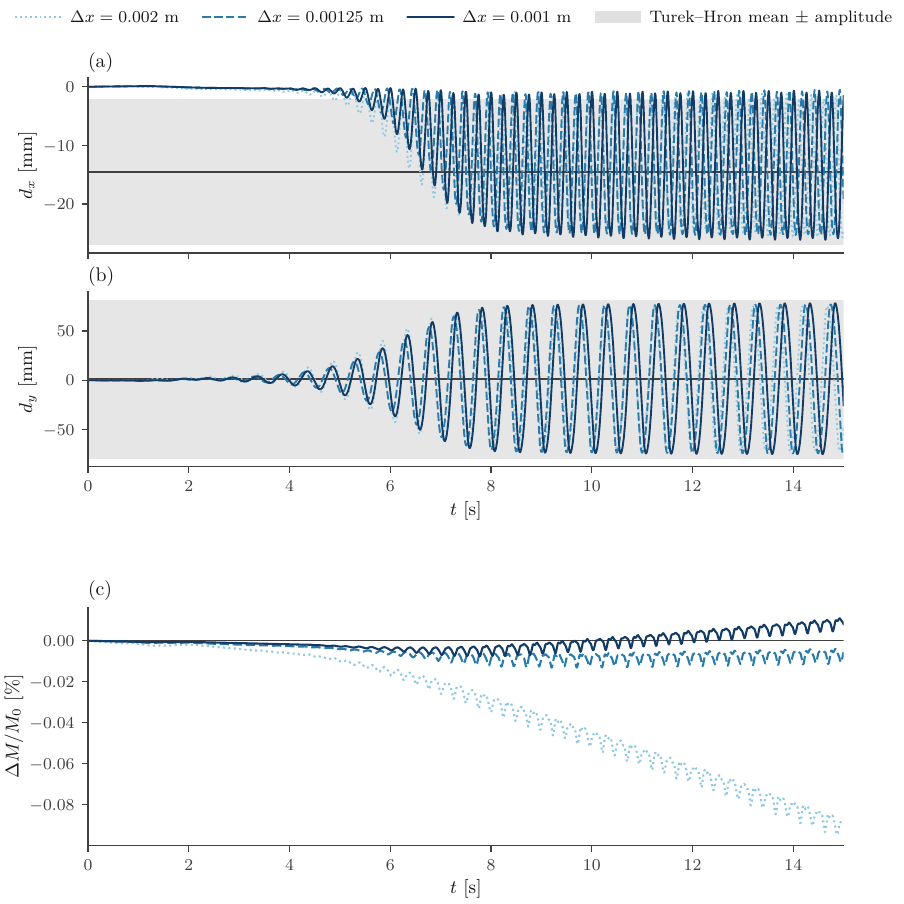}
  \caption{Turek--Hron FSI2, $\beta=0.5$, $\tau_+=0.51$, on three grids:
  (a)~horizontal and (b)~vertical displacement of point $A$, with the range
  mean $\pm$ amplitude (band) and the mean value (line) of the
  reference~\cite{turek2006}; (c)~mass error~\eqref{eq:mass_error} of the
  bar.}
  \label{fig:turek_fsi2}
\end{figure}

\paragraph{FSI2}
The displacement of point $A$ approaches the reference under
refinement (Figure~\ref{fig:turek_fsi2}, Table~\ref{tab:turek_tip}): the frequency
matches the reference from the intermediate grid on, and the deviation of the
vertical amplitude decreases at first order. This is the
$\Order{\varepsilon}$ model error, with $\varepsilon=\Order{\Delta x}$.

\begin{table}[!htbp]
\centering
\caption{Physical parameters of the Turek--Hron FSI2 and FSI3
benchmarks~\cite{turek2006}.}
\label{tab:turek_params}
\begin{tabular}{lcc}
\toprule
Parameter & FSI2 & FSI3 \\
\midrule
$\rho_f$ [$\mathrm{kg\,m^{-3}}$] & $10^3$ & $10^3$ \\
$\nu_f$ [$\mathrm{m^2\,s^{-1}}$] & $10^{-3}$ & $10^{-3}$ \\
$\rho_s/\rho_f$ & 10 & 1 \\
$G_s$ [$\mathrm{Pa}$] & $0.5\times10^6$ & $2.0\times10^6$ \\
$\bar U$ [$\mathrm{m\,s^{-1}}$] & 1 & 2 \\
$Re=\bar UD/\nu_f$ & 100 & 200 \\
\bottomrule
\end{tabular}
\end{table}

\begin{table}[!htbp]
  \centering
  \caption{Turek--Hron benchmarks, $\beta=0.5$ (i.e.\
  $\varepsilon=4\dx$), $\tau_+=0.51$: mean value $\pm$ amplitude and
  frequency of the displacement of point $A$, followed by a material marker,
  over the last five periods, and relative deviation of the vertical
  amplitude and frequency from the reference~\cite{turek2006}.}
  \label{tab:turek_tip}
  \small
  \setlength{\tabcolsep}{3pt}
  \resizebox{\textwidth}{!}{\begin{tabular}{lccccccc}
\toprule
 & & \multicolumn{2}{c}{$d_x$ (point $A$)} & \multicolumn{2}{c}{$d_y$ (point $A$)} & \multicolumn{2}{c}{Deviation from~\cite{turek2006}} \\
\cmidrule(lr){3-4}\cmidrule(lr){5-6}\cmidrule(lr){7-8}
Case & $\Delta x$ [mm] & mean $\pm$ ampl.\ [mm] & $f$ [Hz] & mean $\pm$ ampl.\ [mm] & $f$ [Hz] & ampl.\ $d_y$ & $f_y$  \\
\midrule
FSI3 & $2$ & $-10.26\pm3.10$ & $10.87$ & $+1.06\pm32.78$ & $5.46$ & $-4.7\,\%$ & $+3.1\,\%$  \\
FSI3 & $1.25$ & $-4.09\pm2.81$ & $10.99$ & $+1.22\pm34.76$ & $5.46$ & $+1.1\,\%$ & $+3.1\,\%$  \\
FSI3 & $1$ & $-3.62\pm2.73$ & $10.99$ & $+1.40\pm34.92$ & $5.46$ & $+1.6\,\%$ & $+3.1\,\%$  \\
FSI3 & \multicolumn{1}{l}{Ref.~\cite{turek2006}} & $-2.69\pm2.53$ & $10.90$ & $+1.48\pm34.38$ & $5.30$ & -- & -- \\
\midrule
FSI2 & $2$ & $-13.43\pm12.18$ & $4.07$ & $+1.87\pm73.19$ & $2.04$ & $-9.2\,\%$ & $+2.0\,\%$  \\
FSI2 & $1.25$ & $-12.91\pm12.38$ & $3.99$ & $+1.35\pm75.46$ & $2.00$ & $-6.4\,\%$ & $+0.1\,\%$  \\
FSI2 & $1$ & $-13.37\pm12.75$ & $3.99$ & $+1.56\pm76.47$ & $2.00$ & $-5.1\,\%$ & $-0.2\,\%$  \\
FSI2 & \multicolumn{1}{l}{Ref.~\cite{turek2006}} & $-14.58\pm12.44$ & $3.80$ & $+1.23\pm80.60$ & $2.00$ & -- & -- \\
\bottomrule
\end{tabular}}
\end{table}

\subsection{Discussion of the results}
\label{sec:synthesis}
The observed refinement rates of the benchmarks are rates on the grids
available, not established asymptotic orders, and their weight depends on the
number of grids: the Turek--Hron rates, obtained on three closely spaced
grids, only measure the reduction of the error between them.

The trajectories and the energies converge at first order, as
expected for convergence towards the sharp-interface problem
(Section~\ref{sec:layers}). The mass error then converges at second order
when the jumps across the interface are moderate (lid-driven cavity with
$G_s=0.1$, FSI3), at a slightly lower rate when $\BB$ is strongly stretched
in the layer (Taylor--Green), and at first order when the density jump is
large (Taylor--Green at $\rho_s/\rho_f=10$), a case outside the regime covered
by the analysis (Section~\ref{sec:layers}). In \regII, second order is
recovered for the mass at $\rho_s/\rho_f=10$ (Taylor--Green), consistently with the interfacial contribution
of~\eqref{eq:interface_estimate}. The Taylor--Green
energies converge at first order only in \regII: second order towards the
diffuse-interface solution is established on the oscillating layers, not on
this strongly deforming case. The Jacobian defect, largest in the
interfacial layer, converges at rates between one half and two, depending on
the configuration and on the regime.

\section{Conclusion}
\label{sec:conclusion}

A fully Eulerian lattice Boltzmann method has been developed for
fluid--structure interaction with neo-Hookean solids. The formulation couples
a pressure-based kinetic solver, a polynomial volume-of-fluid reconstruction
with analytically integrated fluxes, and transport of the left Cauchy--Green
tensor on a single Cartesian grid. All updates are explicit and use local
stencils, without a Lagrangian solid mesh, reference-map extrapolation, or a
global pressure solve.

The main contribution is the identification and correction of two defects in
the coupling between the independently transported density and the kinetic
update. A lattice-link correction preserves uniform-translation states
exactly at the discrete level, with round-off errors at density ratios up
to $1000$ (Table~\ref{tab:translation}). A pressure-mode
correction removes the leading frame-dependent contribution that can produce
secular solid-mass drift. Together, these corrections provide a consistent
basis for treating unequal densities while retaining the explicit structure
of the method.

The verification separates the accuracy of the discretisation, second order
towards the diffuse-interface solution at fixed interface thickness, from the
first-order convergence towards the sharp-interface solution when the
thickness scales with the grid spacing. The Turek--Hron benchmarks are
computed in the latter, sharp-interface regime, at density ratios of one and
ten.

The deforming-solid benchmarks presented here cover solid-to-fluid density
ratios up to ten; the larger ratios examined in uniform translation establish
a discrete preservation property rather than general robustness for deforming
solids. The formulation is written in $d$ dimensions and uses only local
operations, but the computations of this paper are two-dimensional; a
three-dimensional implementation is left for future work.

\section*{Funding}

This work was supported by a doctoral contract of the French Ministry of
Higher Education and Research.

\section*{CRediT authorship contribution statement}

\textbf{Tristan Millet:} Conceptualization, Methodology, Software, Validation,
Formal analysis, Investigation, Visualization, Writing -- original draft.
\textbf{Erwan Liberge:} Conceptualization, Methodology, Supervision,
Writing -- review \& editing.

\section*{Declaration of competing interest}

The authors declare that they have no known competing financial interests or
personal relationships that could have appeared to influence the work
reported in this paper.

\section*{Data availability}

Data will be made available on request.

\appendix

\section{MC--MUSCL reconstruction and AB2 integration}
\label{app:muscl}
For a one-dimensional field $b_i$, with $\delta_i^-=b_i-b_{i-1}$ and
$\delta_i^+=b_{i+1}-b_i$, the Monotonized-Central slope is
\begin{equation}
  s_i=\minmod\!\left(2\delta_i^-,\ \tfrac12(\delta_i^-+\delta_i^+),\ 2\delta_i^+\right),
\end{equation}
where $\minmod$ returns the argument of smallest modulus if all have the same
sign and zero otherwise. The face states are
$b_{i+\half}^{L}=b_i+\frac12s_i$ and $b_{i+\half}^{R}=b_{i+1}-\frac12s_{i+1}$.
The advection term is discretised in advective form: the upwind side is
selected by the sign of the nodal velocity, and
\begin{equation}
  \bigl(u\,\partial_xb\bigr)_{ij}
  =\frac{u_{ij}}{\dx}
  \begin{cases}
    b_{i+\half,j}^{L}-b_{i-\half,j}^{L},&u_{ij}\ge0,\\[2pt]
    b_{i+\half,j}^{R}-b_{i-\half,j}^{R},&u_{ij}<0,
  \end{cases}
\end{equation}
with the analogous expression in $y$. The reconstruction is applied to each
component $(B_{xx},B_{xy},B_{yy})$ with independently limited slopes. The
stretching term $\LL\BB+\BB\LL^T$ is evaluated componentwise with
$\LL=\Grad\uu$ from the isotropic operator~\eqref{eq:lbm_grad}, which gives
$\mathcal R^n=\mathcal R(\BB^n,\uu^{n+1})$ in~\eqref{eq:rhs_B}, advanced
by~\eqref{eq:ab2_B}. On the boundary nodes of the domain, $\BB$ keeps its
stress-free value $\Id$ and the slope is set to zero.

\section{Polynomial MTHINC reconstruction}
\label{app:mthinc}

\subsection{Flux reconstruction}
\label{app:mthinc_flux}

\paragraph{Reconstruction and primitives}
In the unit donor cell $(X,Y)\in[0,1]^2$, the reconstruction is
$\mathcal{H}=H\bigl(\beta(n_xX+n_yY+d)\bigr)$, with the cubic smoothed
Heaviside $H(s)=\tfrac12+\tfrac34s-\tfrac14s^3$ for $|s|<1$, $0$ for
$s\le-1$ and $1$ for $s\ge1$. Its primitives $H_1'=H$, $H_2'=H_1$, with
$H_1(-1)=H_2(-1)=0$, are
\begin{equation}
  H_1=\tfrac12 s+\tfrac38 s^2-\tfrac1{16}s^4+\tfrac3{16},
  \qquad
  H_2=\tfrac14 s^2+\tfrac18 s^3-\tfrac1{80}s^5+\tfrac3{16}s+\tfrac1{20}
  \qquad(|s|<1),
\end{equation}
continued by $0$ for $s\le-1$ and by $H_1=s$, $H_2=\tfrac12s^2+\tfrac1{10}$
for $s\ge1$, which makes them $C^1$. The normal $(n_x,n_y)$ is the
normalised Youngs gradient of $\phi$ on the $3\times3$ stencil.

\paragraph{Integral over a rectangle}
Two substitutions $s=\beta(n_xX+n_yY+d)$ give the integral of $\mathcal{H}$
over $[x_0,x_1]\times[y_0,y_1]\subseteq[0,1]^2$ in closed form,
\begin{equation}
  \mathcal{B}(x_0,x_1,y_0,y_1)
  =\frac{H_2(s_{11})-H_2(s_{10})-H_2(s_{01})+H_2(s_{00})}{\beta^2\,n_xn_y},
  \qquad s_{ab}=\beta\bigl(n_xx_a+n_yy_b+d\bigr).
  \label{eq:app_band}
\end{equation}
For $n_y=0$, a single primitive is kept,
$\mathcal{B}=(y_1-y_0)\bigl[H_1(\beta(n_xx_1+d))-H_1(\beta(n_xx_0+d))\bigr]/(\beta n_x)$,
symmetrically for $n_x=0$, and $\mathcal{B}=(x_1-x_0)(y_1-y_0)H(\beta d)$ if
both vanish.

\paragraph{Volume-fraction constraint}
The offset $d$ solves $A(d)\equiv\mathcal{B}(0,1,0,1)=\phi^n_{ij}$. Since
$\partial_dH_2(s_{ab})=\beta H_1(s_{ab})$,
\begin{equation}
  A'(d)=\frac{H_1(s_{11})-H_1(s_{10})-H_1(s_{01})+H_1(s_{00})}{\beta\,n_xn_y}\ge0,
\end{equation}
the integral over the cell of the non-negative kernel $\beta H'$. $A$ is
non-decreasing, constant only where the transition layer lies outside the
cell, and spans $[0,1]$ on
$\bigl[-M-\tfrac1\beta-1,\,-m+\tfrac1\beta+1\bigr]$, with
$m=\min(0,n_x)+\min(0,n_y)$ and $M=\max(0,n_x)+\max(0,n_y)$. For an
interfacial cell, $0<\phi^n_{ij}<1$, the root is therefore unique and is
found by bracketed iteration on this interval.

\paragraph{Flux}
The face velocity is frozen at $t^n$ over the step, an
$\Order{\dx^2}$ error in the diffusive scaling. With
$C=u^n_{i+\half,j}\dt/\dx$, the flux is the integral of the donor
reconstruction over the strip swept upwind of the face,
\begin{equation}
  F^{\phi}_{i+\half,j}=\mathcal{B}(1-C,1,0,1)\ \ (C\ge0),
  \qquad
  F^{\phi}_{i+\half,j}=-\mathcal{B}(0,|C|,0,1)\ \ (C<0),
\end{equation}
the donor being $(i,j)$ or $(i+1,j)$; the $y$-flux follows by exchanging $x$
and $y$. Within a sweep, $\mathbf n$ and $d$ are reused for both faces of
the donor cell, and pure cells use the upwind flux.

\paragraph{Boundedness}
With the dilatation source, the MTHINC flux does not keep $\phi$ in $[0,1]$,
whereas the upwind flux $C\phi_{\mathrm{donor}}$ does: by~\eqref{eq:face_div},
$\dt(\partial_xu)_{ij}=C_{i+\half}-C_{i-\half}$, and each upwind sweep
of~\eqref{eq:thinc_sweep} is a convex combination of its inputs if the
outgoing (implicit sweep) or incoming (explicit sweep) Courant numbers of a
cell sum to at most one. Each sweep is therefore closed by a flux-corrected
transport limiter~\cite{zalesak_fully_1979}: the antidiffusive part of the
MTHINC flux is scaled, face by face, by the largest factor in $[0,1]$ that
keeps both adjacent cells in $[0,1]$, the capacities including the factor
$1-\dt\,\partial_xu$ of the implicit sweep. A single flux per shared face
preserves the discrete conservation of $\phi$.

\subsection{Time integration}
\label{app:mthinc_time}
\label{app:temporal_order}
The reconstruction is rebuilt before each directional
sweep~\eqref{eq:thinc_sweep}. The dilatation source $\phi\Div\uu$ is treated
implicitly (backward Euler) in the first sweep and explicitly (forward Euler)
in the second, and the order and roles are exchanged every step. Each sweep
is first-order in time, but the two forms carry opposite second-order
defects, which cancel at leading order over a two-step cycle; with
$\dt=\Order{\dx^2}$, the splitting error is in any case $\Order{\dx^2}$.

\section{Asymptotic analysis of the pressure-based scheme}
\label{app:asymptotics}

The fully discrete update is expanded directly, following
Junk~\cite{junk_2005}, in lattice units ($c=1$, $\cs^2=1/3$,
$\omega_0=4/9$), in the diffusive scaling and for smooth fields.

\subsection{Diffusive scaling}
\label{app:scaling}
With $\xx=h\boldsymbol\xi$, $t=h^2n$ ($h=\dx$, $h^2=\dt$) and
$D_q=\mathbf{c}_q\cdot\Grad$, one streaming step expands as
\begin{equation}
  e^{hD_q+h^2\partial_t}-1
  =hD_q+h^2\bigl(\partial_t+\tfrac12D_q^2\bigr)
  +h^3\bigl(D_q\partial_t+\tfrac16D_q^3\bigr)
  +h^4\bigl(\tfrac12\partial_t^2+\tfrac12D_q^2\partial_t+\tfrac1{24}D_q^4\bigr)
  +\Order{h^5}.
\end{equation}
The lattice fields scale as $\uu^{\mathrm{LB}}=h\uu+h^3\mathbf{U}$,
$p^{\mathrm{LB}}=h^2p+h^4P$, $\Ftot^{\mathrm{LB}}=h^3\Ftot$ and
$\rho^h=\rho+h^2r$, so that the numerical velocity is
$\uu+h^2\mathbf{U}+\Order{h^4}$, and $f_q=\sum_{k\ge1}h^kf_q^{(k)}$. With
$\psi_q=\omega_q\mathbf{c}_q\cdot\uu/\cs^2$ and
$\chi_q=\omega_q\bigl[(\mathbf{c}_q\cdot\uu)^2/(2\cs^4)-|\uu|^2/(2\cs^2)\bigr]$,
the equilibrium is $f_q^{\mathrm{eq}}=h\rho\psi_q+h^2(a_qp+\rho\chi_q)+\Order{h^3}$, where
\begin{equation}
  \textstyle\sum_q\psi_q=\sum_q\chi_q=0,\quad
  \sum_q\mathbf{c}_q\psi_q=\uu,\quad
  \sum_q\mathbf{c}_q\chi_q=\mathbf{0},\quad
  \sum_q\mathbf{c}_q\mathbf{c}_q\psi_q=\mathbf{0},\quad
  \sum_q\mathbf{c}_q\mathbf{c}_q\chi_q=\uu\uu .
\end{equation}
The forcing is $\mathcal{S}_q=h^3\mathcal{S}_q^{(3)}+h^4\mathcal{S}_q^{(4)}$:
$\mathcal{S}^{(3)}$ is odd and carries $\Ftot$, $\mathcal{S}^{(4)}$ is even,
carries $\uu\Ftot+\Ftot\uu$ and has the moving scalar moment
$\mathcal{S}_M^{(4)}=\sum_{q\neq0}\mathcal{S}_q^{(4)}=\tfrac43\,\uu\cdot\Ftot$.
The linearised collision is $(\mathcal{C}g)_q=\lambda_+g_q^++\lambda_-g_q^-$,
$f^{\mathrm{neq},(k)}=f^{(k)}-f^{\mathrm{eq},(k)}$, and the pressure
closure~\eqref{eq:macro_p} fixes
\begin{equation}
  \textstyle\sum_{q\neq0}f_q^{\mathrm{neq}}=-\frac12\sum_{q\neq0}\mathcal{S}_q .
  \label{eq:app_closure}
\end{equation}

\subsection{Expansion of the link correction and of the regularisation}
\label{app:LC_expansion}
Whatever the density weighting, $\uu_{\ell,q}=\uu+\tfrac h2D_q\uu+\Order{h^2}$,
and $\Delta_q=h^2\Delta_q^{(2)}+h^3\Delta_q^{(3)}+\Order{h^4}$ with
\begin{equation}
  \Delta_q^{(2)}=\psi_qD_q\rho,
  \qquad
  \Delta_q^{(3)}=\psi_q\bigl(\partial_t\rho+\tfrac12D_q^2\rho\bigr)
  +\bigl(\chi_q+\tfrac12D_q\psi_q\bigr)D_q\rho .
  \label{eq:app_Delta}
\end{equation}
$\Delta^{(2)}$ is even and carries no momentum; the momentum contribution
starts with $\Delta^{(3)}$. The regularisation increment $\mathcal{Q}_q$
satisfies, at every order,
\begin{equation}
  \textstyle\sum_{q\neq0}\mathcal{Q}_q=0,\quad
  \sum_q\mathbf{c}_q\mathcal{Q}_q=\mathbf{0},\quad
  (\Pi_{xx}-\Pi_{yy})[\mathcal{Q}]=\Pi_{xy}[\mathcal{Q}]=0,\quad
  \mathcal{Q}_0=0,
  \label{eq:app_Q_invariants}
\end{equation}
and acts only on the two odd ghost modes and the isotropic even mode; since
$f^{(1)}=\rho\psi_q$ lies in the retained subspace, $\mathcal{Q}^{(1)}=0$.

\subsection{Orders \texorpdfstring{$h$}{h} and \texorpdfstring{$h^2$}{h2}: incompressibility and viscous stress}
\label{app:order12}
At order $h^2$,
\begin{equation}
  D_q(\rho\psi_q)=-(\mathcal{C}f^{\mathrm{neq},(2)})_q+\Delta_q^{(2)}+\mathcal{Q}_q^{(2)} .
  \label{eq:app_order2}
\end{equation}
Summing over $q\neq0$ with~\eqref{eq:app_closure}
and~\eqref{eq:app_Q_invariants} gives $\Div{\rho\uu}=\uu\cdot\Grad\rho$,
i.e.\ $\Div\uu=0$; without the link correction one would obtain
$\Div{\rho\uu}=0$, incompatible with the transport of $\rho$. Since
$D_q(\rho\psi_q)=\psi_qD_q\rho+\rho D_q\psi_q$, the density gradient cancels
in~\eqref{eq:app_order2}; the remaining source $\rho D_q\psi_q$ is pair-even,
momentum-free, and its scalar moment and trace vanish with $\Div\uu$. It lies
in the range of $\mathcal{P}$, so $\mathcal{Q}^{(2)}=0$ and
\begin{equation}
  f_q^{\mathrm{neq},(2)}=-\frac{\rho}{\lambda_+}D_q\psi_q,
  \qquad
  \textstyle\sum_q\mathbf{c}_q\mathbf{c}_qf_q^{\mathrm{neq},(2)}
  =-\frac{\rho\cs^2}{\lambda_+}\bigl(\Grad\uu+\Grad\uu^T\bigr):
  \label{eq:app_neq2}
\end{equation}
the regularisation changes neither the leading stress nor the viscosity.

\subsection{Order \texorpdfstring{$h^3$}{h3}: momentum equation}
\label{app:order3}
At order $h^3$,
\begin{equation}
  \partial_tf_q^{(1)}+D_qf_q^{(2)}+\tfrac12D_q^2f_q^{(1)}
  =-(\mathcal{C}f^{\mathrm{neq},(3)})_q
  +\bigl[(I-\tfrac12\mathcal{C})\mathcal{S}^{(3)}\bigr]_q
  +\Delta_q^{(3)}+\mathcal{Q}_q^{(3)} .
\end{equation}
In the first moment, $\mathcal{Q}$ does not contribute, and
$\sum_q\mathbf{c}_qf_q^{\mathrm{neq},(3)}=-\tfrac12\Ftot$
from~\eqref{eq:macro_u} reduces collision and forcing to $\Ftot$. The link
terms $\psi_q\partial_t\rho+\chi_qD_q\rho$ balance the term
$\uu(\partial_t\rho+\uu\cdot\Grad\rho)$ of
$\partial_t(\rho\uu)+\Div{\rho\uu\otimes\uu}$, and
$\tfrac12\psi_qD_q^2\rho+\tfrac12D_q\psi_qD_q\rho$ combine with
$\tfrac12D_q^2(\rho\psi_q)$ into $\tfrac12D_q(\rho D_q\psi_q)$, of first moment
$\tfrac12\cs^2\Div{\rho(\Grad\uu+\Grad\uu^T)}$. With~\eqref{eq:app_neq2},
\begin{equation}
  \rho\Bigl(\frac{\partial\uu}{\partial t}+\uu\cdot\Grad\uu\Bigr)
  =-\Grad p+\Div{2\rho\nu\DD(\uu)}+\Ftot,
  \qquad
  \nu=\cs^2\Bigl(\frac1{\lambda_+}-\frac12\Bigr),
  \label{eq:app_NS}
\end{equation}
equivalent to the conservative form~\eqref{eq:full_mom} when
$\partial_t\rho+\uu\cdot\Grad\rho=0$ at leading order.

\subsection{Order \texorpdfstring{$h^4$}{h4}: pressure--divergence relation}
\label{app:order4}
At uniform density, the equilibrium moving moment is
$m^{(2)}=\tfrac53p+\tfrac43\rho K$, $K=\tfrac12|\uu|^2$. By~\eqref{eq:app_closure},
collision and forcing contribute
$-\lambda_+(-\tfrac12\mathcal{S}_M^{(4)})+(1-\tfrac{\lambda_+}2)\mathcal{S}_M^{(4)}
=\tfrac43\uu\cdot\Ftot$ to the moving scalar equation, independently of
$\lambda_+$:
\begin{equation}
  \partial_tm^{(2)}+\Div{\mathbf{j}^{(3)}}+\tfrac12\Grad\Grad:\Pi^{(2)}
  +\partial_t\Div{\mathbf{j}^{(1)}}+\tfrac16\Grad\Grad\Grad\,\vdots\,Q^{(1)}
  =\tfrac43\,\uu\cdot\Ftot,
\end{equation}
with $\mathbf{j}^{(k)}$, $\Pi^{(k)}$, $Q^{(k)}$ the first three moments of
$f^{(k)}$; $\mathcal{Q}$ adds no term. On the incompressible branch, the
$\mathbf{j}^{(1)}$ and $Q^{(1)}$ terms vanish,
$\mathbf{j}^{(3)}=\rho\mathbf{U}-\tfrac12\Ftot$, and
$\Grad\Grad:\Pi^{(2)}=\Div\Ftot$ cancels $-\tfrac12\Div\Ftot$, so that
$\tfrac53\partial_tp+\tfrac43\rho\,\partial_tK+\rho\Div{\mathbf{U}}=\tfrac43\uu\cdot\Ftot$.
Eliminating $\rho\,\partial_tK$ with $\uu\cdot$\eqref{eq:app_NS},
\begin{equation}
  \frac{Dp}{Dt}+\frac35\rho\Div{\mathbf{U}}=\mathcal{N}_p,
  \label{eq:app_material}
\end{equation}
with $\mathcal{N}_p$ from~\eqref{eq:Np}; the work of $\Ftot$ cancels thanks to
the second-order forcing and its term in~\eqref{eq:macro_p}. Where $\rho$
varies smoothly, the link correction with the mass-weighted link velocity and
a uniform $\nu$ adds $\mathcal{N}_\rho$, \eqref{eq:Nrho}, to the right-hand
side; the modified equation of the phase transport adds further,
discretisation-dependent terms, not specified here.

\subsection{Mass of the solid at second order}
\label{app:mass}
Consider a solid that does not touch $\partial\Omega$. The phase
transport~\eqref{eq:phi_conservative} is a material transport by the
numerical velocity $\mathbf{U}$, whose divergence follows
from~\eqref{eq:app_material}:
$\Div{\mathbf{U}}=-\kappa(\rho)\bigl(Dp/Dt-\mathcal{N}_p-\mathcal{N}_\rho\bigr)$,
$\kappa(\rho)=(1-\omega_0)/(\rho\cs^2)$. With $W=\phi\,\kappa(\rho(\phi))$,
the kinetic contribution to the mass $M=\int_\Omega\phi$ is
\begin{equation}
  M(t)-M(0)
  =h^2\Bigl[
   \underbrace{-\!\int_0^t\!\!\int_\Omega W\frac{Dp}{Dt}}_{\text{material}}
   \underbrace{+\!\int_0^t\!\!\int_\Omega W\mathcal{N}_p}_{\text{frame defect}}
   \underbrace{+\!\int_0^t\!\!\int_\Omega W\mathcal{N}_\rho}_{\text{density gradient}}
  \Bigr]+\Order{h^4};
  \label{eq:app_mass}
\end{equation}
the phase transport and, in \regI, the interfacial layer add their own
errors~\eqref{eq:interface_estimate}. The three terms behave differently:
\begin{itemize}[leftmargin=*,itemsep=1pt]
  \item the material term equals $-[P(t)-P(0)]$, $P=\int_\Omega Wp$, since
  $DW/Dt=0$: an end-point term, bounded by the pressure;
  \item the density-gradient term equals
  $-\tfrac{3\cs^2}{20}[E_\rho(t)-E_\rho(0)]$, with
  $E_\rho=\int_\Omega(W/\rho)|\Grad\rho|^2\ge0$, which follows the length of
  the interface: also an end-point term;
  \item the frame defect is not a time derivative: for a rigid translation at
  velocity $\mathbf{V}$, it drifts at the constant rate
  $-\tfrac95\kappa\,\mathbf{V}\cdot\mathbf{F}_p$, where $\mathbf{F}_p$ is the
  pressure force on the solid, and vanishes in the frame of the solid. The
  projection~\eqref{eq:pressure_projection_main}, for which
  $\delta_p^{\mathrm{LB}}=h^4\mathcal{N}_p+\Order{h^6}$, removes it.
\end{itemize}
After the projection, the mass therefore varies only by bounded end-point
terms. At uniform density, $h^2\kappa$ is the compressibility
$\dt^2/(\rho c_a^2\dx^2)$ and the material term reads
$\Delta M/M=-Ma_{\mathrm{el}}^2\,\Delta\langle p\rangle_\phi/G_s$, with
$\langle p\rangle_\phi$ the mean pressure in the solid: the kinetic mass error
scales as $\dx^2$ at fixed $\tau_+$ and as $(\tau_+-\tfrac12)^2$ on a fixed
grid.

\bibliographystyle{elsarticle-num}
\bibliography{citations}
\end{document}